# From stars to diverse mantles, melts, crusts and atmospheres of rocky exoplanets

**Claire Marie Guimond[1], Haiyang Wang[2], Fabian Seidler[2], Paolo Sossi[2], Aprajit Mahajan[3], and Oliver Shorttle[4,5,*]**

[1]Atmospheric, Oceanic, and Planetary Physics, University of Oxford, UK

[2]Institute of Geochemistry and Petrology, ETH Zurich, Clausiusstrasse 25, CH-8092, Zürich, Switzerland

[3]Trinity College, University of Cambridge, UK

[4]Department of Earth Sciences, University of Cambridge, UK

[5]Institute of Astronomy, University of Cambridge, UK

[*]Corresponding author: shorttle@ast.cam.ac.uk

## 1 Introduction

The advent of radial velocity (RV) and transit photometry method has resulted in the detection of more than 5000 exoplanets since 1995. Of these detected planets, a significant fraction, ~20%,[1] may be dominantly rocky. Given the biases inherent to exoplanet detection, this implies an even larger fraction of the overall planet population as likely possessing a significant rock fraction (e.g., Fulton et al., 2017). However, owing to the difficulty in detecting Earth-sized planets, the vast majority of these rocky planets are much larger than Earth (so-called 'super-Earths'); with radii $\lesssim$ 1.5 times that of Earth ($1.5R_E$) and with masses 2–8 $M_E$ (Earth masses; Fulton and Petigura, 2018). Because the radial velocity technique yields estimates of planetary masses, and the transit method permits measurements of their radii, together, they can be used to infer bulk densities (e.g., Charbonneau et al., 1999; Valencia et al., 2007). The uncompressed densities of such super-Earths are similar to the terrestrial planets of our Solar System (Figure 1).

To first order, similarity to solar system terrestrial planet densities may imply that the compositions and core/mantle ratios of super-Earths are also similar (e.g., Dorn et al., 2015; Santos et al., 2015; Dressing et al., 2015). However, a planet's estimated density is often degenerate with respect to its composition, and the same density

---

may be consistent with many combinations of plausible planet-forming materials (Valencia et al., 2007; Rogers and Seager, 2010): In particular, there is typically uncertainty over what the relative contribution to a planet's overall

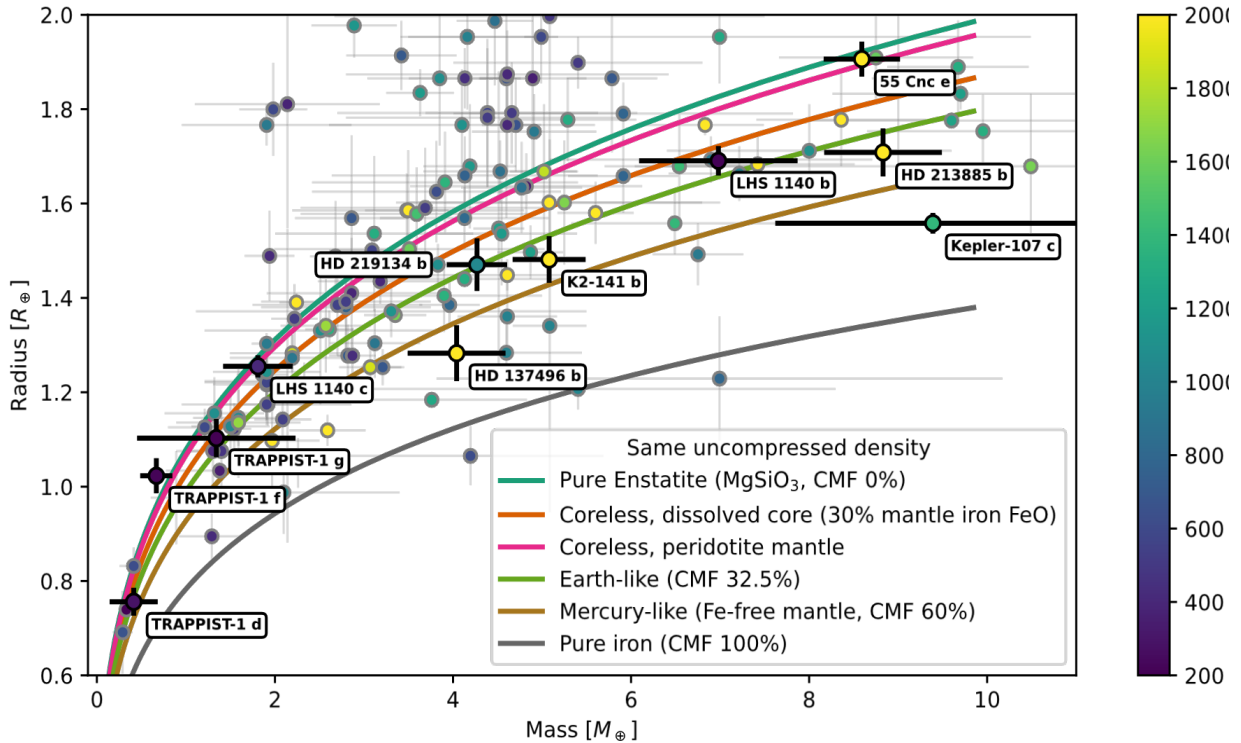

**Figure 1.** Mass-radius diagram for existing low-mass exoplanets, colored by their calculated equilibrium temperatures. The different curves represent mass-radius trends for various interior properties. The exoplanet data were obtained from https://exoplanet.eu/catalog/; only planets with masses and radii with uncertainty less than 70% are shown, and some interesting candidates are highlighted by name. The mantle compositions for the interior models are obtained from McDonough and Sun (1995) (coreless peridotite & Earth-like), Nittler et al. (2018) (Mercury) and Wang et al. (2022a) (coreless, dissolved core).

mass and radius is of dense rock/metal interior compared with low density atmosphere/envelope. Even with the knowledge that a planet has no or very little atmosphere (e.g., Kreidberg et al., 2019), degeneracies remain between the mantle to core ratio and with composition from mass-radius measurement alone (e.g., Unterborn et al., 2016; Dorn and Lichtenberg, 2021). In the solar system context the polar moment of a inertia of a planet can help elucidate the mass distribution at depth; whilst for exoplanets this property of a planet may in future be inferred, it is likely to be extremely challenging to apply with useful accuracy in the case of small rocky planets (being more readily applied to fluid planets, e.g., Padovan et al., 2018; Consorzi et al., 2023). As a result of this uncertainty on planetary composition, the use of stellar elemental abundances (measured with stellar photospheric spectroscopy), have been proposed to further constrain the bulk compositions, and by extension interior structures, of rocky exoplanets on both individual and population levels (cf. Sotin et al., 2007; Grasset et al., 2009; Delgado Mena et al., 2010; Thiabaud et al., 2015; Santos et al., 2015; Dorn et al., 2017; Hinkel and Unterborn, 2018; Wang et al., 2019a; Putirka and Rarick, 2019; Schulze et al., 2021; Wang et al., 2022a,b; Spaargaren et al., 2023; Unterborn et al., 2023a)

How the measured stellar abundances translate to those in bulk planets remains uncertain. Progress has been made by applying basic truisms from our understanding of the Solar System that are assumed to hold for exoplanetary systems; namely, that the refractory elements (of which Ca, Al and Ti are the most abundant) are present in rocky planets in roughly solar (stellar) proportions (e.g., Ringwood, 1966; Wanke et al., 1974). Although the major, rock-forming¨ elements, Mg, Si, and Fe behave in a more volatile manner than refractory Ca-Al-Ti (e.g., Lodders, 2003), they are also near-solar in proportions in the Earth compared to the Sun (Palme et al., 2014; McDonough and Yoshizaki, 2021, near-solar, but with important differences; ). This provides the basis for estimating (refractory) rock-forming elemental compositions in rocky planets from their measured host stellar compositions for such elements. More uncertain are the abundances of volatile elements, and O in particular (Wang et al., 2019b), the depletion of which—dubbed "volatility trend"—is shown to vary across the solar system rocky bodies (relative to the Sun; Larimer and Anders, 1967; Wasson and Chou, 1974; Anders and Ebihara, 1982; Bland et al., 2005; Moynier et al., 2011; Braukmuller et al., 2019; Wang¨ et al., 2019b; Fegley et al., 2020; McDonough and Yoshizaki, 2021; Khan et al., 2022) and is modulated by (often unknown) star-planet early interactions and planet formation history (Albarede, 2009; Hin et al., 2017; Norris and` Wood, 2017; Sossi et al., 2019; Bitsch and Battistini, 2020). Therefore, on top of assuming a stellar composition starting point, empirical constraints from solar system rocky planet compositions have been additionally proposed to constrain the bulk compositions of (rocky) exoplanets for both refractories and volatiles (Wang et al., 2019a, 2022a,b; Spaargaren et al., 2023). The simulation of physical processes during planet formation, particularly nebular condensation, have also been performed to predict exoplanetary compositions (Bond et al., 2010b,a; Moriarty et al., 2014; Dorn et al., 2019; Jorge et al., 2022; Timmermann et al., 2023).

Of all the major planet-forming elements, oxygen stands out as having first-order importance. The O budget in a rocky planet is central in determining two fundamental planetary properties, its core mass fraction and mantle Fe content: properties which, if known, would help resolving degeneracies in relating exoplanet densities to bulk compositions. Putirka and Rarick (2019) treat the planetary Fe/FeO ratio as a free parameter, an approach also adopted by Unterborn et al. (2023b) and Guimond et al. (2023b), to predict the likely compositional range of rocky exoplanets (treated as a free parameter, because simply in cosmochemical terms there is almost always enough oxygen available to fully oxidise Fe and other rock forming elements, so oxygen abundances must be forced lower in planets than stellar abundances allow for cores to form at all Putirka and Rarick, 2019). On this basis, these authors conclude that the mantle compositions of rocky exoplanets bear a strong resemblance to that of the Earth, with major differences being limited to Mg/Si ratios. However, uncertainty in planetary Fe/FeO ratios, together with the stochastic nature of the accretionary behaviour of volatile elements (e.g., Albarede, 2009; Sossi et al., 2022) mean that the assumptions underpinning our use of the` star-planet connection to model exoplanet compositions remains observationally untested: atmospheres (if present) trade off against interior mineralogy and structure (core mass fraction) to make density alone a weak constraint on planetary composition (e.g., Rogers and Seager, 2010).

Consequently, it is, as yet, unknown whether the Earth-like uncompressed densities of super-Earths speak to a similarity in composition, or whether significant chemical variation exists among the population. Resolving this question is important for the field, as it has the potential to speak to a key result that has emerged from exoplanet surveys over the last decade: that there is a drop in planet occurrence rates at $\sim 1.5R_E$ (Fulton et al., 2017). This feature of the data has been linked to envelope loss from sub-Neptune-sized planets initially formed in this radius interval— a loss with several hypothesised physical drivers, e.g. photoevaporation (Owen and Wu, 2013, 2017) and core-powered mass loss (Ginzburg et al., 2018; Gupta and Schlichting, 2019). How similar the composition of the remaining envelope-free kernels of these sub-Neptunes are to 'born rocky' planets remains uncertain. Improved constraints on planet bulk compositions and interior structures may offer the possibility of distinguishing these two types of small ($< 1.5R_E$) planets from one other; illuminating a dichotomous evolutionary path for small planets that we do not have an analogue for in the solar system.

Attempts to 'cut the Gordian knot' with respect to the exoplanetary compositions have used empirical trends between planets and stellar abundance data. One such trend indicates that the excess density of super-Earths scales positively with the fraction of iron in the star (Adibekyan et al., 2021), suggesting a genetic connection between the two. However, the range in densities among the terrestrial planets spans almost the entire range observed in exoplanets ($\leq 1.5R_E$) within error (Figure 1), such that the compositions of rocky exoplanets remain equivocal from this approach. There are however hints at planetary compositions and/or structures not represented by the solar system population, in particular those planets resolvably less dense than Earth or Mercury (Figure 1)—although whether this is due to the contribution of atmospheres/envelopes or interior structure (such as the absence of a core) remains to be seen.

The next generation of observations are therefore needed to better constrain planetary compositions. These observations are likely to be manifest in two forms; *i)* detection of key gaseous species in the atmospheres of exoplanets and, *ii)* characterisation of surface features of (potentially) bare rocky exoplanets (Kreidberg et al., 2019; Greene et al., 2023). Because the detectability of such transmission/emission spectral features in the infrared scales proportionally with the temperature and radius of the planet, the most promising targets are those rocky super-Earths that are very close to their host star and have large, extended atmospheres and/or transparent atmospheres that facilitate observations of the surface — surfaces which may be molten due to their intense heating from their star.

This review is focused on describing the logic by which we make predictions of exoplanetary compositions and mineralogies, and how these processes could lead to compositional diversity among rocky exoplanets. We use these predictions to determine the sensitivity of present-day and future observations to detecting compositional differences between rocky exoplanets and the four terrestrial planets. First, we review data on stellar abundances and infer how changes in composition may manifest themselves in the expected bulk compositions of rocky exoplanets (section 2). Converting this information in mass-radius relationships requires calculation of the stable mineral assemblages at a given temperature-pressure-composition ($T$−$P$−$X$), an exercise we describe in section 3. Should the planet be hot enough to engender partial melting of the mantle, then these liquids are likely to rise to the surface and erupt to form planetary crusts; the possible compositional and mineralogical variability of which we examine in section 4. Finally, the expected spectroscopic responses of such crusts are examined in section 5. This approach, taking us from star to planet, is summarised as a flow chart in Figure 2.

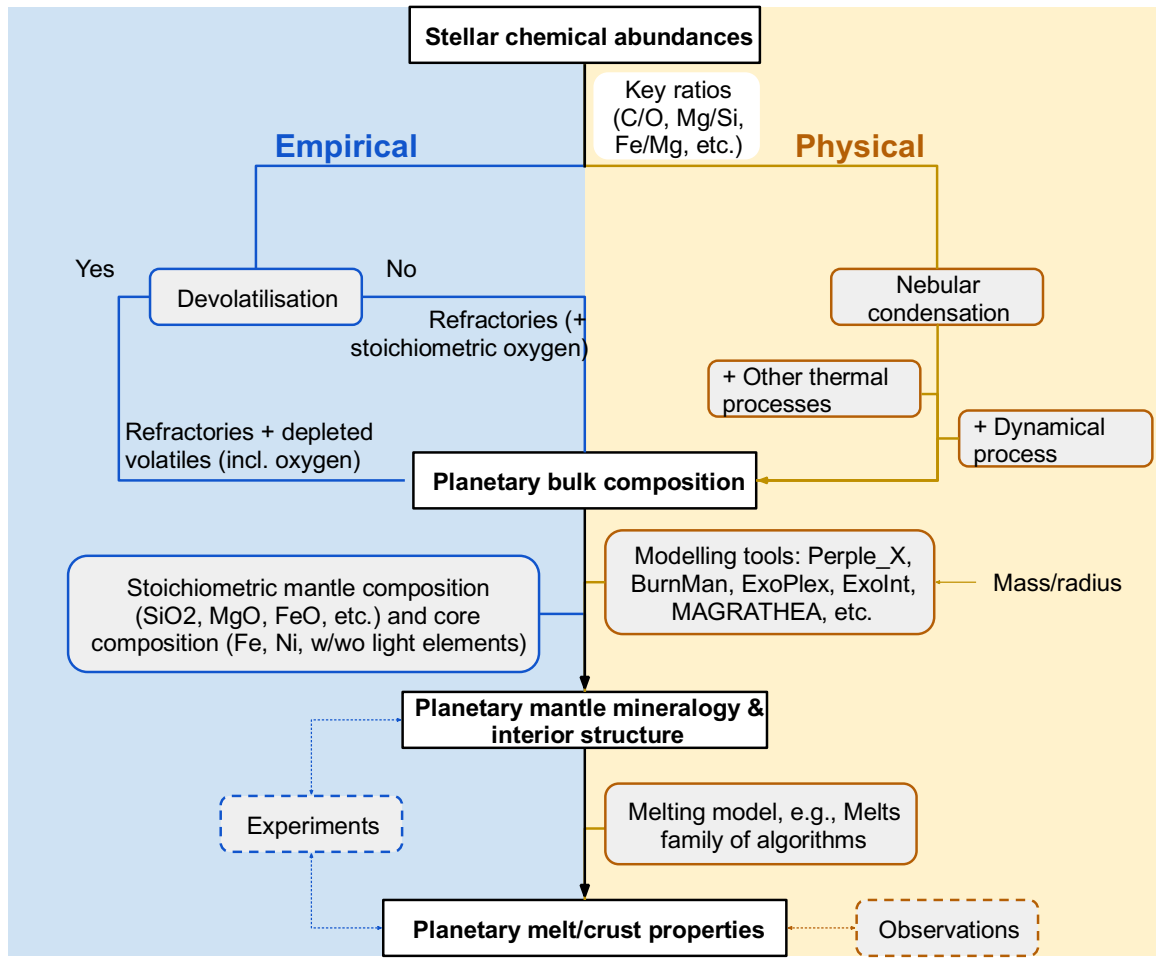

**Figure 2.** The outline of the review from stars to mantles, melts and crusts and from models to experiments and observations.

## 2 Bulk chemical compositions of rocky exoplanets

### 2.1 The planet-star connection

Planets and their host stars are born from the same molecular cloud material, giving them a shared chemical origin. Gas-dust separation in the protoplanetary disk and subsequent dynamical evolution may cause the bulk chemical compositions of (particularly rocky) planets to deviate from that of their host star. However, by analogy with the composition of the Earth's mantle and that of the Sun, the *relative proportions* of refractory (rock-forming) elements between rocky exoplanets and their host stars are expected to be within ~10 % relative (Palme et al., 2014; Wang et al., 2018). This assumption has been carried over into hypothetical exoplanetary systems (Dubois et al., 2002; Sotin et al., 2007; Schulze et al., 2021, e.g.,). The veracity of this assumption is largely untested (with the exception of weak support coming from polluted white dwarf observations, Bonsor et al., 2021), and, in the case of non-refractory, major elements (Mg, Si and Fe), may lead to large discrepancies (>10%) between true compositions and those predicted based on the assumption that these elements are present in stellar proportions (Dauphas et al., 2015; Wang et al., 2019a; Miyazaki and Korenaga, 2020). This is because the Mg/Si ratio, for example, varies by a factor ~2 even among chondritic meteorites (Wasson and Kallemeyn, 1988). Nevertheless, the advantage of adopting stellar abundances as proxies for those of potential surrounding planets is that they are measurable with stellar photospheric spectroscopy (Huang et al., 2005; Hinkel et al., 2014; Nissen et al., 2017; Delgado Mena et al., 2017; Buder et al., 2018; Liu et al., 2020; Adibekyan et al., 2021; Hourihane et al., 2023) and so provide a broad observationally-informed bound on planetary composition.

Stellar chemical compositions vary across the galaxy because stars have formed in different locations at different times and inherit the local chemical history of stellar birth, nucleosynthesis, and death (Gaidos, 2000; Bond et al., 2008; Mojzsis, 2022; Pignatari et al., 2023). That is, the compositions of stars are inherited from the molecular cloud from which they formed. Abundance ratios of elements with respect to H are typically reported in 'dex' (i.e., base 10 logarithmic) units: typical ranges for FGK stars in stellar surveys fall between -0.5 and +0.5 dex of the solar value;

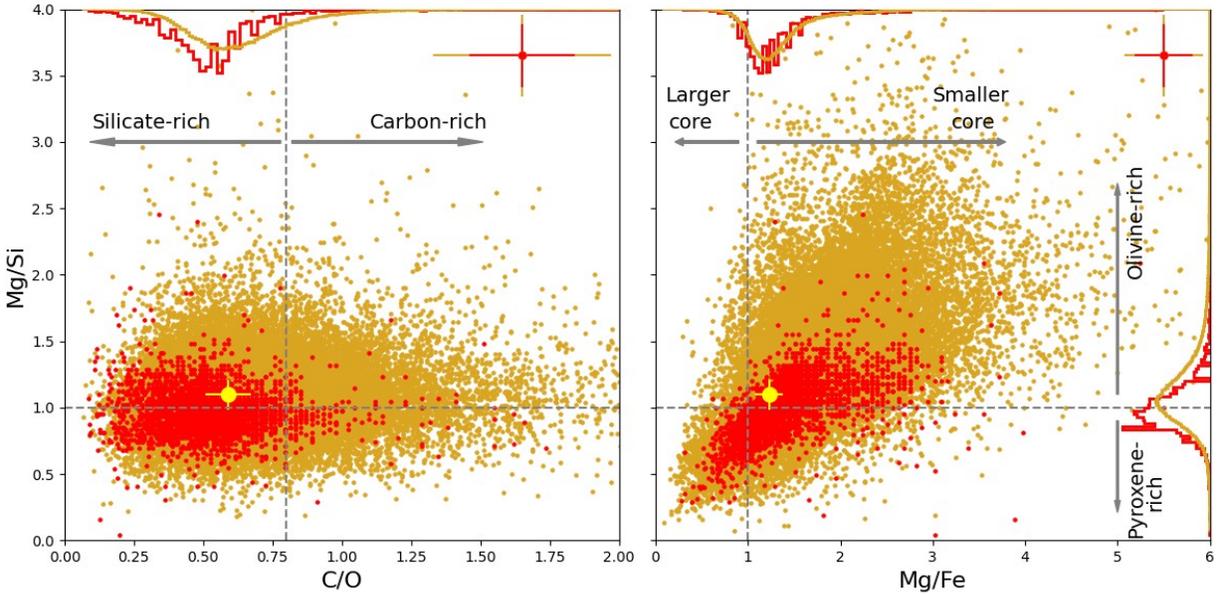

**Figure 3.** Key elemental ratios – C/O, Mg/Si and Mg/Fe – of FGK stars in using both GALAH (yellow; Buder et al. 2021) and Hypatia (red; Hinkel et al. 2014) stellar catalogues. The histograms of these ratios in both databases are also shown, along with the typical error bars of these ratios. The reference solar abundances (in bright yellow), which have been used to normalise both databases, are from Asplund et al. (2021). The apparent fewer number of stars in the higher Mg/Si regime (> 2.0) in the left panel (compared to the right panel) is due to the fewer available measurements of carbon and oxygen abundances in the GALAH database. The vertical and horizontal dash lines (C/O:0.8, Mg/Fe:1.0, and Mg/Si:1.0), along with different arrows, indicate how composition may lead to mineralogical and structural propensities of hypothetical rocky exoplanets around these stars.

measurement errors can be as low as 0.01 dex as reported by individual groups, but are more likely to be on the order of 0.1 dex between groups given the non-standardised techniques for quantifying abundances from stellar spectra (Hinkel et al., 2016; Hinkel and Unterborn, 2018). Some key elemental ratios in stars—namely, C/O, Mg/Si and Mg/Fe—have been proposed to indicate potential properties of rocky exoplanets around those stars (Bond et al., 2010a,b; Thiabaud et al., 2015; Wang et al., 2019a). Figure 3 shows the spread of these key elemental ratios for FKG stars in the solar neighbourhood in using two different stellar catalogues (GALAH, Buder et al. 2021; Hypatia, Hinkel et al. 2014). The Sun appears to be rather typical in both catalogues (aligning well with the medians of the corresponding histograms). These key ratios indicate, to first order, the propensity of mantle composition and core sizes of hypothetical rocky exoplanets around these stars (see indications in Figure 3). The large spreads in these parameter spaces reveal the potential chemical diversity of rocky worlds around such stars.

A diversity of host star elemental abundances implies a diversity in planetary bulk compositions (e.g., Sotin et al., 2007; Delgado Mena et al., 2010; Bond et al., 2010b; Carter-Bond et al., 2012a; Hazen et al., 2015; Unterborn and Panero, 2017; Hinkel and Unterborn, 2018; Putirka and Rarick, 2019; Putirka et al., 2021; Wang et al., 2022a; Mojzsis, 2022; Guimond et al., 2023a,b; Spaargaren et al., 2023). However, because the abundances of the major rock-forming elements are positively correlated in FGK stars, the resultant ratios of these elements vary by only modest amounts (factor 2–3; e.g., Hinkel et al., 2014; Young et al., 2014; Brewer et al., 2016). Consequently, should elemental abundances in a given star be a good proxy for that of the planet, then this would imply that rocky exoplanets are composed predominantly of O, Fe, Mg, Si, together with smaller amounts of the refractory elements Ca and Al – largely concordant with our experience of solar system planet composition and mineralogy (Putirka et al., 2021).

## 2.2 Nebular condensation

The current paradigm for the mechanisms by which the terrestrial planets formed involves a stage of nebular cooling and condensation. Collection of these nebular condensate grains into pebbles, prior to their mixing and growth into progressively more massive bodies, dubbed planetesimals ($< 10^{-3} M_E$) and embryos ($\sim 0.1 M_E$), is thought to have led to the formation of the solar system's four inner planets (e.g., Morbidelli et al., 2012; Johansen and Dorn, 2022).

The composition of the solar nebular gas is closely approximated by that of the Sun (Urey, 1954, Larimer and Anders, 1967; Wasson and Chou, 1974; Anders and Ebihara, 1982; Alexander et al., 2001; Bland et al., 2005; Moynier et al., 2011; Hin et al., 2017; Norris and Wood, 2017; Wang et al., 2019b; Sossi et al., 2019; Braukmuller et al., 2019; Fegley¨ et al., 2020). Yet, this process of condensation is not isochemical, as the observed volatile abundances in solar system rocky bodies (e.g., C, O, S, P, K and Na) are depleted compared to their abundances in the Sun, whereas refractory lithophile elements are relatively enriched: in the bulk silicate Earth, this factor of refractory lithophile enrichment is 1.21 × solar, normalised to Mg (Palme et al., 2014)—normalising to the refractory lithophile element Al results in an identical baseline for refractories between the terrestrial planets and the Sun (Wang et al., 2019b; McDonough and Yoshizaki, 2021; Sossi et al., 2022).

The main piece of (empirical) evidence pointing to the importance of nebular condensation in dictating the compositions of the terrestrial planets is that their abundances of volatile elements decrease as a smooth function of their 50 % nebular condensation temperatures ($T_c^{50}$; Fig. 4). This is defined as the temperature at which 50 % the mass of an element is condensed from a solar composition gas at $10^{-4}$ bar (e.g., Larimer and Anders, 1967; Lodders, 2003; Wood et al., 2019). Such a trend holds among chondritic meteorites, although they exhibit constant-abundance plateaus in the most volatile elements (with $T_c^{50} < 700$ K; Braukmuller et al. 2019). In this manner, rocky planets are, by definition,¨ those that have lost (or never acquired) the major complement of the volatile elements present in the Sun. For example, H is depleted by a factor $\sim 10^6$ in the Earth (relative to the Sun, normalised by Al), equating to $\sim 1000\ \mu g/g$ (Marty, 2012; Hirschmann and Withers, 2008). Only O, S, C, and H are likely to be sufficiently abundant to affect mass-radius relationships in an observable manner. Though their abundances in rocky exoplanets have been yet to be measured, they must, by analogy, be depleted to similar levels to those observed in the Earth (with the possible exception of C, see below), lest the mass-radius relationships of rocky exoplanets diverge from the observed, Earth-like trend (e.g., Luque and Palle, 2022).

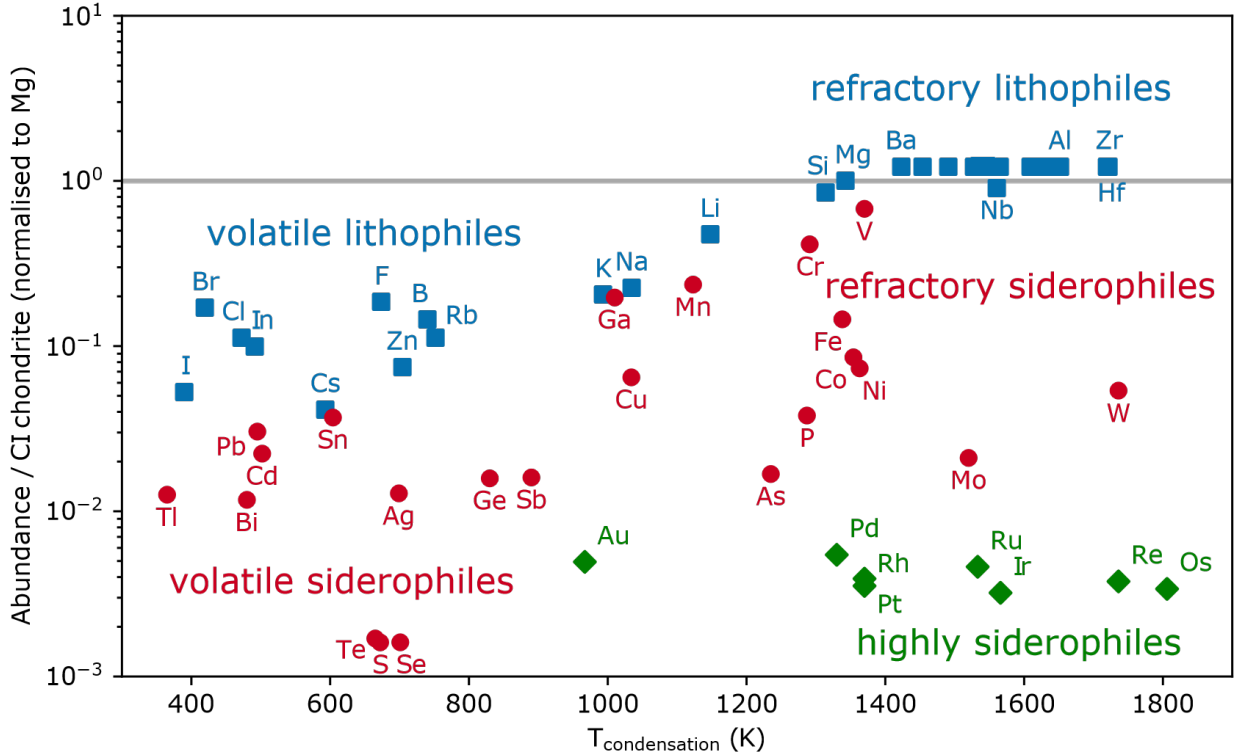

**Figure 4.** An example of the volatility trend of elements in rocky objects in the solar system, in this case the Earth. The Earth's composition has been normalised to CI chondrites, a model for the presumptive building blocks of rocky material in the solar system, were they not to have variably lost volatile elements. Normalisation of the Earth and CI composition to Mg accounts for the large hyper-volatile component within chondrites. Compositional data are from Palme et al. (2014) and Clay et al. (2017), half mass condensation temperatures are from Wood et al. (2019).

The difficulty in predicting the volatile budgets of rocky exoplanets lies in the complexity of the processes that lead to their accretion. Were the volatile element abundances of Earth to have resulted from nebular condensation alone, then elements that condense below a given threshold temperature, $T_{cut-off}$ should be absent from the solid material from which the Earth accreted. By contrast, the Earth should harbour its full (solar) complement of elements with $T_c^{50} > T_{cut-off}$. This is because elements typically condense over a narrow range of temperatures, passing from fully vaporised to fully condensed over a temperature range of ~30–40 K (e.g., Larimer, 1967; Albarede, 2009). That the Earth exhibits a smooth decline in volatile element abundances with $T_c^{50}$ and not a binary pattern with an inflection point at some fictive $T_{cut-off}$, indicates that these elements were brought through mixing of different materials that each experienced various $T_{cut-off}$ (Sossi et al. 2022). As a result, planetary compositions cannot be simply tied to condensation temperature, and, instead, are a complex function of their accretion histories. This problem has been circumvented by applying 'devolatilisation trends' defined for the Earth (and other telluric bodies) that quantify element abundance *vs.* $T_c$[50] to provide limits as to the expected volatile element abundances in rocky exoplanets (Wang et al., 2019a, 2022a,b; Spaargaren et al., 2023).

It is noteworthy that there is no requirement for the condensation temperatures of a nebula around another star to conform to those for a solar composition gas. Indeed, variations in element ratios, particularly C/O (Larimer and Bartholomay, 1979) may lead to drastic differences in condensation temperatures, brought about by a change in the condensing mineral assemblage. For example, carbides and sulfides (such as oldhamite, a common phase in enstatite chondrites) become stable at the expense of oxides (cf. Larimer and Bartholomay, 1979; Bond et al., 2010b) for nebulae with C/O ratios higher than $\sim 0.9$. The corollary of this result is that elements that are thought of as being volatile in a Solar System context (namely C) behave as refractory elements (and vice-versa). This realisation has led to some experimental work done on reduced phases, such as SiC, which is supposed to dominate rocky exoplanets with high C/O ratios (Miozzi et al., 2018; Hakim et al., 2018; Hakim et al., 2019; Allen-Sutter et al., 2020). As such, exotic compositions beyond silicate/pyrolitic mantles may be plausible (see section 3.2).

## 2.3 The role of O in controlling core/mantle ratios

Among the volatile elements, oxygen is expected to be by far the largest constituent of planetary mantles, owing to its abundance in stellar nebulae and the fact that it combines readily with metals to form oxides and silicates. As a result, O plays a critical role in modulating the composition of the mantle and thus the stable mineral phases therein. This is because the finite amount of O that condenses from stellar nebulae combines sequentially with the rock-forming oxides according to their redox potentials (at 1 bar, this order is Ca > Al > Mg > Si > Fe; Chase et al. 1982), until it is exhausted. Typically, O/(Al+Ca+Mg+Si+Fe) ratios in condensing phases are such that some fraction of Fe is left over as metallic iron ($Fe^0$). Such a system with $Fe^0$ and FeO present and in equilibrium defines the oxygen fugacity ($fO_2$; i.e., the partial pressure of $O_2$ gas corrected for non-ideality) via:

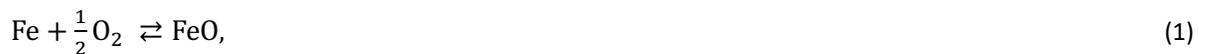

$$Fe + \frac{1}{2}O_2 \rightleftarrows FeO, \tag{1}$$

where higher proportions of FeO/Fe indicate more oxidising conditions. The reduced iron, together with other 'reducing' (or siderophile) metals (e.g., Ni and Co, and notably Si, which imparts mineralogical consequences discussed later) constitute the core of the planet, and hence determine the core mass (Wang et al., 2019a, 2022a).

Because oxygen condenses gradually from stellar nebulae as a function of temperature, there is, as yet, no satisfactory parameterisation that would permit its abundance in planet-forming materials to be predicted *a priori*. One insight is that varying the Mg/Si ratio of the nebula would alter the olivine/pyroxene ratio (which have (Mg,Fe)/O = 1/2 and (Mg,Fe)/O = 1/3, respectively) and thereby the amount of oxygen consumed during formation of the silicate portion of a planet (Unterborn and Panero, 2017). However, this is only one of the routes via which oxygen may be incorporated into planets. Therefore, oxygen abundance is typically presumed to be sufficient to oxidise major silicate-building elements until Fe (i.e., Mg, Si, Ca, Al; Santos et al. 2015; Dorn et al. 2015; Brugger et al. 2017; Plotnykov and Valencia 2020; Schulze et al. 2020; Guimond et al. 2023b). This assumption is loosely supported by the observation that O/(Mg + Si + Fe + Ca + Al) is sufficiently high across the vast majority of Hypatia Catalog stars (Putirka and Rarick, 2019). Meanwhile, to account for variable condensed oxygen budgets, the degree to which metals like Fe (and Ni) are oxidised is typically prescribed: as a fixed value (nil or at Earth's value; Santos et al. 2015; Schulze et al. 2020; Spaargaren et al. 2023); or, as a nominal range by introducing an additional parameter, such as mantle Mg or Fe number (Mg# = $\frac{Mg}{Mg+Fe}$; Fe# = $\frac{MFeg}{Mg+Fe}$; Sotin et al. 2007; Brugger et al. 2017; Plotnykov and Valencia 2020, or Fe partition coefficient (Putirka and Rarick, 2019; Guimond et al., 2023b; Unterborn et al., 2023b), with bounds chosen to imply Fe-metal cores ubiquitously. As a result, whilst host stellar abundances may approximate *bulk* rocky planetary compositions, there is debate over how these can be used to place constraints on the interior structure and properties (i.e., densities) of rocky planets (Wang et al., 2019a; Schulze et al., 2020; Seidler et al., 2021; Adibekyan et al., 2021; Unterborn et al., 2023b; Spaargaren et al., 2023).

**2.4 Post-accretionary processes**

Processes within planets themselves compound the challenge of linking stellar composition through to internal structure. A key process supposed to be operative early in a planet's history is Fe disproportionation, where a high pressure reaction of the form

$$3FeO \rightleftarrows Fe^{0} + Fe_2O_3, \tag{2}$$

causes ferrous iron, $Fe^{2+}$, hosted in the silicate portion of the planet, to break down into $Fe^{3+}$ and metallic iron (Frost et al. 2004; Armstrong et al. 2019). The metal, if it sinks to the core, will result in net oxidation of the complementary mantle with no change in the amount of O in the bulk planet. Other processes, such as the dissociation of $H_2O$ and loss of $H_2$ by atmospheric escape, followed by mantle influx of the oxygen, may lead to a similar effect on the mantle oxygen budget (Catling et al., 2001; Zahnle et al., 2019; Wordsworth et al., 2018; Krissansen-Totton et al., 2021, see also Hirschmann 2023).

FeO may also be produced during core formation by the dissolution of Si into metal,

$$SiO_2^{silicate} + 2Fe^0 \rightleftarrows 2FeO^{silicate} + Si^0, \tag{3}$$

occurring at high temperatures and pressures relevant for the base of a deep magma ocean (Javoy, 1995). Reaction (3) leads to still more mantle oxidation. This reaction also proves important by increasing the Mg/Si ratio of the mantle with respect to the bulk planet, a complexity which will be returned to in section 3.

These post-accretionary processes are significant for the subsequent coupled interior-atmosphere evolution of a planet. There must have been a reset of how oxidising Earth's mantle was following core formation (where it would have been highly reducing; e.g., Righter and Ghiorso, 2012) to its present day more oxidising state (Frost and McCammon, 2008). Iron disproportionation reactions like (2), whether driven by mineral growth or in the silicate liquid itself (Hirschmann, 2022), presumably contributed to this evolution. The result is a mantle ultimately producing much more oxidising volcanic gases than it would have done immediately following core formation. Such post-accretionary shifts in mantle redox have potentially profound implications for resultant atmospheres and oxygen cycles in the planet (Ortenzi et al., 2020; Liggins et al., 2022), yet the effects of which remain hard to quantitatively predict. Improved observational constraints on the composition of Venusian magmas and volcanic gasses would provide an important new data point from the solar system on how these processes and others play out on roughly Earth-sized planets.

## 3 Mantle mineralogy of rocky exoplanets

The major rocky reservoirs on planets are their mantles. Initially, prior to the formation of thick crusts, this mantle is a crystallised rocky layer that has the composition of the bulk planet, less those elements that partitioned into its metallic core. The stable assemblages of minerals within these mantles vary across planets according to their bulk compositions, as this section will detail. The resulting mineralogies control planetary thermochemical evolution insofar as different minerals have different material properties (e.g., density, bulk moduli and heat capacity; Spaargaren et al., 2020). Across the common mantle minerals and assemblages thereof, notable differences are suggested in, for example: viscosities (e.g., Yamazaki et al., 2009; Ammann et al., 2011; Hansen and Warren, 2015; Girard et al., 2016; Thielmann et al., 2020; Cordier et al., 2023), volatile storage (see section 3.3; e.g., Dong et al., 2022; Wang et al., 2022b; Guimond et al., 2023b), solidii (e.g., Hirschmann, 2000; Kiefer et al., 2015; Brugman et al., 2021), and ferric iron partitioning and thus upper mantle oxygen fugacity (Guimond et al., 2023a). In these ways, predicting rocky planet mineralogies represents an emerging frontier in exoplanet characterisation.

The mineralogy of a planetary mantle is set by equilibria established among its constituent phases. The stability of a given (mineral) phase assemblage for a fixed bulk composition is dictated by the minimisation of Gibbs free energy at a given $P{-}T$ (e.g., Connolly and Kerrick, 1987). Pressure and temperature profiles through planets are predicted using equations of state and adiabatic gradients, calculated iteratively with mineralogy. The bulk composition may be estimated from host star abundances, though this inevitably introduces uncertainty: first, due to poorly-constrained partitioning processes, as detailed in section 2; and second, due to measurement error on the stellar abundances themselves. Uncertainties of 0.2 dex on two elements propagate to a fractional uncertainty on their molar ratio of 65%, for example (using the derivation in Hinkel et al., 2022). A further, underlying assumption is that this bulk oxide composition is homogeneous throughout the mantle; there is some debate about whether this is true for Earth (e.g., Anderson and Bass, 1986; Ballmer et al., 2017).

Having estimates of bulk composition, mantle phase equilibria can then be modelled self-consistently with pressuretemperature profiles, most commonly using thermodynamic algorithms that minimise Gibbs free energy, a class of models including Perple X (Connolly, 2009) and HeFESTo (Stixrude and Lithgow-Bertelloni, 2011). The user of these algorithms must further commit to a thermodynamic database and choice of activity composition relations for the expected minerals (e.g., Ghiorso and Sack, 1995; Holland and Powell, 2011; Green et al., 2016; Stixrude and Lithgow-Bertelloni, 2022). Because such databases and mineral models are experimentally constrained for primarily Earth-like compositions, they necessarily require a degree of extrapolation to the exoplanetary context: this extrapolation is both in terms of composition, where, e.g., Na contents, Mg/Fe ratios, and oxygen fugacity may differ from the terrestrial mantle, and in pressure, where in super-Earth's pressures in the lower mantle may exceed 600GPa, or five times that of Earth's lower mantle (Wagner et al., 2012; Unterborn and Panero, 2019; Boujibar et al., 2020).

As a result of the above, estimates of the mineralogical diversity present in exoplanetary interiors, especially as planets become much more massive than Earth, need cautioning by the limitations of our thermodynamic models. As one example, the oft-used equation of state of crystalline Mg-postperovskite is experimentally validated to 265GPa (Sakai et al., 2016), although higher-pressure measurements have been achieved using shock compression (Spaulding et al., 2012; Bolis et al., 2016; Fratanduono et al., 2018)—recent interior models tend to rely on *ab initio* simulations by Sakai et al. (2016) that estimate the $MgSiO_3$ equation of state to well beyond the plausible rocky exoplanet mantle pressures. Recognising these limitations and assumption, we proceed in this section to explore exoplanet mineralogy.

### 3.1 Possible range of mantle mineralogies for silicate planets

In the past half-decade, a gamut of theoretical studies have exploited phase equilibria models to estimate exoplanet mantle mineralogy from stellar abundances: Hinkel and Unterborn (2018) and Wang et al. (2022a) on nearby stars with higher-precision abundances; and Putirka and Rarick (2019), Unterborn et al. (2023b), Spaargaren et al. (2023), and Guimond et al. (2023b) on a population level of stars from the Hypatia Catalog (Hinkel et al., 2014) and/or the GALAH survey (Buder et al., 2018). A study of mantle compositions inferred from polluted white dwarfs provides a complementary approach (Putirka and Xu, 2021). These studies suggest that rocky planet mantles can, for the most part, be understood in terms of known terrestrial mineral components: i.e., familiar minerals are stabilised in these mantles, even if their relative proportions and composition differ from those in Earth's mantle.

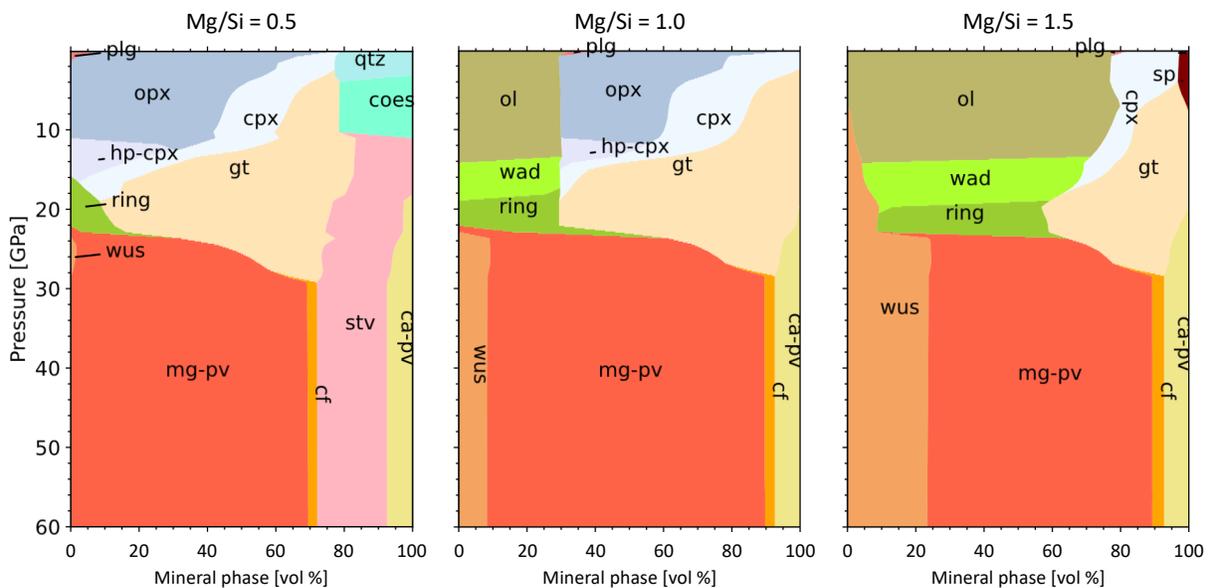

**Figure 5.** The effect of Mg/Si on the mineralogy of a silicate mantle. Except for Mg and Si, the compositions of other elements are kept the same as those of bulk silicate Earth (McDonough and Sun, 1995). The mantle potential temperature is assumed at 1700 K. This is modelled with pyExoInt (Wang et al., 2022a) that has Perple X (Connolly, 2009) integrated for mineralogy output. The abbreviations of mineral phases stand for: ol - olivine, opx - orthopyroxene, cpx - clinopyroxene, hp-cpx - high-pressure clinopyroxene, plg - plagioclase, qtz - quartz, coes - coesite, sp -spinel, wad - wadsleyite, ring - ringwoodite, gt - garnet, stv - stishovite, wus - magnesiowustite (ferropericlase),¨ mg-pv - magnesium perovskite (bridgmanite), cf - calcium ferrite, and ca-pv - calcium perovskite.

#### 3.1.1 Upper mantles

As Mg and Si are the most abundance oxide-forming elements synthesized in stars, the ferromagnesian silicates olivine (and its polymorphs; wadsleyite and ringwoodite) $(Mg,Fe)_2SiO_4$ and orthopyroxene $(Mg,Fe)SiO_3$ are the dominant minerals at pressures where they are stable, <25GPa, depending on temperature. As pressure increases, clino- and orthopyroxene dissolve into garnet to form majorite (Akaogi and Akimoto, 1977), which becomes predominant near ~ 15 GPa, and also contains the entirety of the Ca and Al budget of the planet at these pressures (Figure 5). That is, no observed stellar composition appears rich enough in what would be minor oxides in Earth's mantle (e.g., CaO, $Al_2O_3$, $TiO_2$) to saturate their pure phases (Putirka and Rarick, 2019). Indeed, as shown in Figure 6, 89% of Hypatia stars lead to upper mantles dominated by a mixture of olivine plus orthopyroxene, assuming for simplicity $Mg/Si_{mantle}$ = $Mg/Si_{star}$ and 90 mol% of planets' Fe in their cores; or 85% of stars assuming instead a constant 5 wt% Si in metal cores (e.g., Hirose et al., 2021).

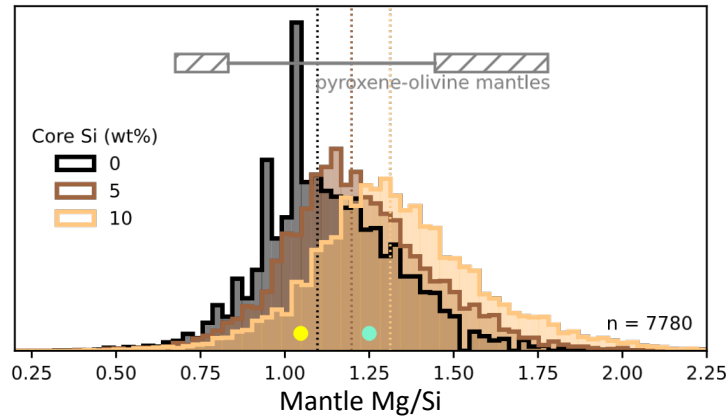

**Figure 6.** The distribution of mantle Mg/Si implied from stellar abundances in the Hypatia Catalog (Hinkel et al., 2014), for different assumptions about the Si content of metal cores (assumed constant for simplicity), and no devolatilisation. Calculations assume 85 mol% of the bulk planet's Fe forms a core; reasonable changes do not affect the distributions. Vertical dotted lines show the median of each distribution. The yellow marker indicates the solar value from Lodders et al. (2009); the blue marker indicates the Bulk Silicate Earth estimate. The horizontal bar shows where mantles are dominated by a mixture of olivine and pyroxene, with the hatched regions showing where only some compositions stabilise both these phases. Calculations based on the Stixrude and Lithgow-Bertelloni (2022) thermodynamic database implemented in Perple X (version 6.9).

Substantial diversity is nonetheless expected within these pyroxene-olivine compositions, even with relative abundances in host stars not significantly different from the sun's. Again, due to their abundance, the most important changes are predicted in the ratio of olivine to orthopyroxene, being a direct function of Mg/Si (Ringwood, 1989; Bond et al., 2010b; Delgado Mena et al., 2010; Young et al., 2014; Hinkel and Unterborn, 2018; Putirka and Rarick, 2019; Wang et al., 2022a,b; Mojzsis, 2022; Spaargaren et al., 2023; Guimond et al., 2023b). Increasing Mg/Si above unity produces more forsterite olivine and the expense of enstatite orthopyroxene—forsterite's chemical formula dictates two units MgO per unit $SiO_2$, versus 1:1 in enstatite (Figure 3, right panel). The extreme ends of the stellar Mg/Si distribution see silicate mantles becoming olivine-free at Mg/Si $\lesssim$ 0.8 (in the presence of some mantle FeO), or orthopyroxene-free at Mg/Si $\gtrsim$ 1.6 (Figure 6). In these cases, the oxides in excess ($SiO_2$ or $(Mg,Fe)O$ wüstite) could¨ form their own phases—one hypothetical avenue for "exotic" mineralogies (section 3.2).

The third-most abundant oxide in most mantles, FeO, has a secondary role in mineralogy, mostly via stabilising olivine polymorphs. Higher Fe/Si would be associated with increased ferrosilite olivine over fayalite orthopyroxene, an analogous effect to increased Mg/Si favouring forsterite. In fact, the mantle ratio (Mg + Fe)/Si is a better predictor than Mg/Si of the ratio olivine/orthopyroxene, yet it is in practice perhaps less useful because of the challenges predicting the Fe content in an exoplanet mantle (see previous section). In any case, the thermodynamic models quickly become unreliable at very FeO-rich compositions ($\gtrsim$ 25wt%); such mantles' mineralogies have not been studied in detail.

Lastly, higher $Al_2O_3$ or CaO proportions would be accommodated by increased garnet and clinopyroxene, up to a maximum at 4GPa of about one third garnet by mass. Components beyond Mg-Si-Fe-Al-Ca are not expected to occur in abundances high enough to notably affect mantle mineralogy—but they may affect other important material properties—solidus temperatures, for example—as section 4 will discuss.

### 3.1.2 Lower mantles

As pressures increase within a planet, several key phase transitions lead to a layered mantle structure. Ringwoodite, the high-pressure polymorph of olivine, breaks down together with majorite, to form bridgmanite (magnesium perovskite) and ferropericlase (magnesiowüstite) (See Chapter 6, Table X, of this issue for a comprehensive list of mineral formulae). This transition occurs at ~22–27 GPa in the Earth's mantle, depending on temperature, and defines the lower mantle (e.g., Ito and Takahashi, 1989; Shim et al., 2001; Dong et al., 2021a, also see Figure 5). Extrapolation of the Stixrude and Lithgow-Bertelloni (2011, 2022) thermodynamic databases suggests that the mantles of most rocky exoplanets, provided that they are comprised largely of silicates and reach sufficient pressures, will have a transition zone-like structure.

At the high pressures of lower mantles (see Figure 5), there is typically less information available about the variation in phases and phase proportions with depth than in upper mantles, so mineralogies tend to be simplified. $(Mg,Fe)SiO_3$ perovskite and postperovskite always dominate here, with a smaller amount of $(Mg,Fe)O$ wüstite (i.e., ferropericlase)¨ taking up any excess Mg and Fe. $CaSiO_3$ perovskite (davemaoite), Earth's third-most-common lower mantle phase, also appears as the main high-pressure host of Ca across the stellar-abundance-derived range of bulk compositions. Most Al is expected to be dissolved in perovskite (Caracas and Cohen, 2005), although some models include Al in a low-abundance calcium ferrite solution as well (Stixrude and Lithgow-Bertelloni, 2022). The main axis for mineralogical diversity in the lower mantle is the effect of increasing Mg/Si on increasing ferropericlase modality at the expense of perovskite (Figure 5). Finally, perovskite will transition to postperovskite around ~120GPa along Earth-like pressure-temperature profiles (Tsuchiya et al., 2004; Murakami et al., 2004; Oganov and Ono, 2004). Because the lower mantle will dwarf the upper mantle in mass for all rocky planets of at least Earth's mass, we expect the prototypical exoplanet mantle to be mostly perovskite and post-perovskite (Figure 7).

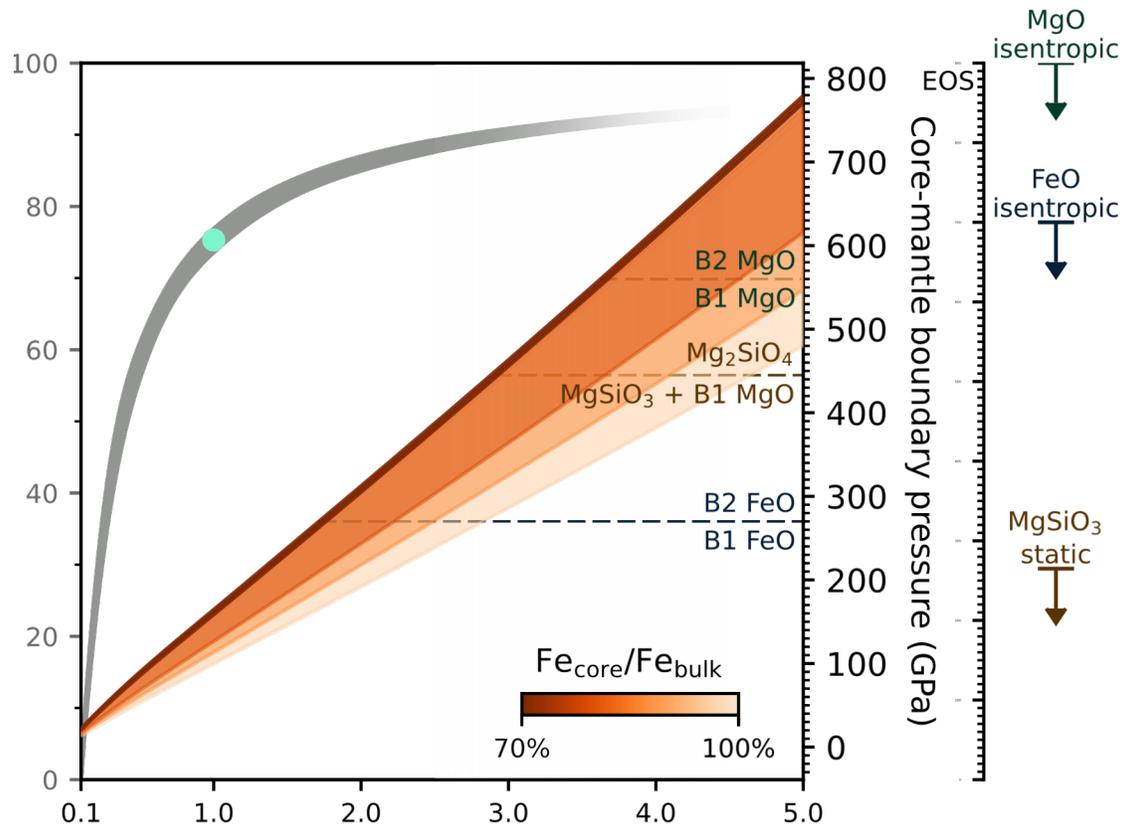

**Figure 7.** *(Left:)* The mass fraction of the mantle where perovskite (pv) or postperovskite (ppv) is stable (grey curve) increases rapidly with increasing planet mass. This mass fraction represents the lower mantle of the planet, and does not vary appreciably with Fe partitioning into the core, potential temperatures between 1600–1900 K, or diverse silicate compositions—the line width spans the 1$\sigma$ distribution over these three variables. Earth's position on the left y-axis is shown with a blue marker. The right y-axis, for context, corresponds to the scaling of core/mantle boundary pressure with planet mass, with lines coloured by fixed partitioning of Fe into the core. Line widths span the 2$\sigma$ distribution over Hypatia-derived Fe/Mg and Si/Mg, and with core Si content varying from 0 to 10% (and no other core light elements for simplicity), and at a mantle potential temperature of 1700K. All values of Fe$_{core}$/Fe$_{bulk}$ result in a similar upper limit to core/mantle boundary pressure, but a lower limit decreasing with increasing Fe in the core, due to the core/mantle boundary moving outward in radius. These core/mantle boundary pressure calculations use the ExoPlex mass-radius composition calculator (see Unterborn et al., 2023b, for details), which implements a liquid Fe core, and the Stixrude and Lithgow-Bertelloni (2011) database of silicate thermodynamic properties in Perple X, assuming as a caveat that all equations of state are valid at the relevant T-P-X conditions. Dashed lines indicate expected endmember phase transitions, not necessarily included in current interior models: in FeO and MgO as observed in ramp-compression experiments from Coppari et al. (2013, 2021), and in MgSiO$_3$ as theoretically predicted by Umemoto et al. (2017, their lower limit with Mg/Si = 2). *(Right:)* Experimental limits to pressure-density equations of state. MgSiO$_3$ is from static-compression experiments by Sakai et al. (2016), and MgO and FeO are again from Coppari et al. (2013, 2021). Note that shock-compression experiments, measuring irreversible compression which may not represent equilibrium states, have reached higher pressures (e.g., Spaulding et al., 2012; Root et al., 2015; Bolis et al., 2016; Sekine et al., 2016; Fratanduono et al., 2018), but are not included here.

The existence of exoplanets more massive than Earth motivates extending silicate phase equilibria to much higher pressures (Figure 7). Whereas Earth's core/mantle boundary (CMB) lies at 130GPa, the highest mantle pressures reached in "rocky" planets, $R_p < 1.5R_E$, were calculated by Unterborn and Panero (2019) to be over five-fold greater, ~630–740GPa, increasing with mantle FeO content (the higher value corresponds to Mars-like mantle FeO; this assumes pure solid $\varepsilon$-Fe cores and Mg/Si = 1). At these extreme lower mantle pressures, silicates may in fact be entirely molten, even at the 'cooler' geotherms of planets several billion years old (Stixrude, 2014; Fratanduono et al., 2018; Boujibar et al., 2020). Enduring basal magma oceans are thus a ubiquitous possibility, presenting an additional challenge for interior structure modellers (and yet to be understood consequences for planetary evolution).

'Super-Earth' pressures are nonetheless accessible to shock experiments, coupled with *in-situ* synchrotron techniques, to probe states of matter at extreme conditions. One of the key mineralogical transformations occurring at high pressure is the transition from high- to low-spin iron (Badro, 2014; Shim et al., 2023). Spin-transitions in Fe may result in significant changes in the elastic properties of exoplanetary mantles, particularly in oxidising, FeO-rich mantles. Further, changes in FeO and MgO crystal lattice structures have been detected in dynamic compression experiments (Coppari et al., 2013, 2021, Figure 7). These and other possible deep-mantle phase transitions are highly relevant for mantle dynamics, via their negative, endothermic Clapeyron slopes: slowing down convection by consuming latent heat (Christensen and Yuen, 1985).

Molecular dynamics simulations represent an alternative route to study the deep interiors of exoplanets. These simulations suggest that at extremely high pressures silicates will dissociate into their constituent oxides (MgO, FeO, $SiO_2$) (Umemoto et al., 2006; Tsuchiya and Tsuchiya, 2011; Umemoto and Wentzcovitch, 2011; Niu et al., 2015; Umemoto et al., 2017). Before complete dissociation, however, Umemoto et al. (2017) found $MgSiO_3$ postperovskite to recombine with either MgO or $SiO_2$ to form new phases starting at ~450–600GPa, depending on the Mg/Si ratio. For example, the oxygen-excess phase $Mg_2SiO_5$ was predicted beyond 600 GPa. As reviewed in Duffy and Smith (2019), limitations of both theoretical and experimental approaches mean it will be important to reconcile results from both, if we are to constrain the mineralogy of deep exoplanet mantles.

The uncertainties on the equations of state and phase diagrams of silicate minerals at ultrahigh pressure may not be the primary source of error when interpreting mass-radius relationships of exoplanets (Unterborn and Panero, 2019). Yet, understanding minerals' associated material properties at these conditions will be key to linking interior dynamics with surface observeables beyond just mass and radius. Doing so will help us answer, for example, how vigorously a deep super-Earth mantle is convecting, if it is convecting at all (e.g., Karato, 2011; Stamenkovic et al., 2011, 2012;´ Tackley et al., 2013; Miyagoshi et al., 2014, 2018; Ritterbex et al., 2018; van den Berg et al., 2019; Coppari et al., 2021; Shahnas and Pysklywec, 2021); and whether its likely mode of convection would help or hinder recycling at the planet surface.

### 3.2 Beyond Earth-like mantles

The foregoing discussion has made the implicit assumption that mantles of exoplanets are similar to those of the four terrestrial planets; that is, that they are predominantly composed of silicates. However, other mineralogies may occur, and even our solar system points to one alternate path for "rocky" planet mantles: the interior of Mercury is purported to contain an FeS layer sitting at its point of neutral buoyancy, between the core and the mantle (Malavergne et al., 2010, though see Cartier et al. 2020). A key variable in driving non-Earth-like mantles is how reducing the planet formation environment was. For disks with C/O>0.9, Condensation sequences may diverge markedly from that of our Solar System (see section 2; Bond et al., 2010b; Moriarty et al., 2014) leading to potentially non-silicate-based rocky exoplanets. However, observations thus far suggest that such high-C/O systems are rare ($\sim$ 1%) among FGK dwarfs in the solar neighbourhood (Teske et al., 2013; Brewer et al., 2016).

For C/O ratios higher than $\sim$0.9, condensation calculations predict carbides and nitrides, rather than oxides and silicates, condense from cooling protoplanetary disk gas (Larimer and Bartholomay, 1979). Planets forming in such settings would be correspondingly enriched in carbon and be highly reducing. As a consequence, these planets should contain iron exclusively in its metallic form (e.g., as $Fe^0$ in their cores rather than as FeO in their mantles). Titanium carbide (TiC) may be an abundant minor phase in the mantles of such planets, whereas other, typically lithophile elements (Mg and Ca) may occur predominantly as sulfides; niningerite and oldhamite (Skinner and Luce, 1971). Planets made from such material are therefore richer in C and S and poorer in O and Al with respect to terrestrial planets. Indeed, such mineralogies have been invoked, in part, to explain the low bulk density of some planets (e.g., Madhusudhan et al., 2012). Coupled chemical-dynamical planet formation simulations of Carter-Bond et al. (2012b) suggest that it is indeed possible to achieve nearly 50% C by mass for inner planets in the upper end of high-C/O disks.

At these highly reducing conditions ($fO_2$ between 5 and 6.5 log units below the IW buffer, and between 2 and 10GPa) Schmidt et al. (2014) indicate that moissanite (SiC) will be stable, and silicate phases will have Mg # between 0.993 and 0.998 (i.e., almost entirely Fe-free). More reducing still, at nine log units below IW, Si metal is stable, however, this is likely to be a lower bound, as Si will dissolve into Fe metal (Lacaze and Sundman, 1991). At higher $fO_2$ than ΔIW-5, SiC may break down to form FeSi liquids and graphite (Hakim et al., 2018, 2019). However, a comprehensive experimental investigation of the stabilities of C- and S-rich exoplanet interiors is currently lacking.

Even among "silicate" compositions, the observed distribution of stellar Mg/Si could also imply the existence of extreme mineralogies that do not fit a pyroxenite or peridotite classification (Putirka and Rarick, 2019). Indeed, the actual distribution of mantle Mg/Si across rocky exoplanets may be wider than the distribution seen in stars. Although the Earth has a superchondritic Mg/Si ratio, as inferred from peridotites of the Earth's upper mantle, yielding Mg/Si of 1.27 (by mass), there are other materials in the Solar System with markedly lower Mg/Si. These materials are represented by the enstatite chondrites, with Mg/Si = 0.64 (Wasson and Kallemeyn, 1988) and therefore contain enstatite rather than forsteritic olivine as the predominant silicate phase.

Extrapolating even further might suggest exoplanetary mantles could become silica ($SiO_2$)-saturated. This condition is likely rare for two main reasons. First, Mg and Si are both major elements in the stellar nebulae, and are not readily fractionated from one another during condensation: This is because they condense into olivine and orthopyroxene, and their condensation reactions are similarly dependent on $fO_2$. Second, Si is more siderophile than is Mg (e.g., Ringwood and Hibberson, 1991; Fischer et al., 2015) and therefore core formation will invariably result in an increase in the Mg/Si ratio of the planetary mantle relative to its bulk composition.

In summary, the major changes occurring to planetary composition, in addition to stellar compositional variability itself and where 0.3 < C/O < 0.9, take the form of:

- Changes in oxidation state during accretion and the magma ocean stage (i.e., whilst the planet's growing metal core is still open to mass exchange with its mantle), which influence, in particular, the Fe/FeO and, under increasingly reducing conditions, $Si/SiO_2$ ratio of the planet, and hence ultimately the mass ratio of core to mantle.

- Fractionation of major silicate forming components Mg and Si from Fe during chemical processing in the protoplanetary disk (Dauphas et al., 2015; Miyazaki and Korenaga, 2020). This may result from physical separation of their main mineral hosts olivine, orthopyroxene and Fe metal, during condensation of the nebular gas and have given rise to their variable abundances in chondrites.

- The extent of volatile element depletion, either during condensation or subsequent accretion and evolution. Owing to their high bulk densities, the class of exoplanets discussed here are not expected to contain substantial quantities ($\gg$1 wt%) of water (as $H_2$ or $H_2O$). Fractional vaporisation and loss of the atmospheres of planets close to their host star may also lead to changes in the chemical composition of the residual planet, because vaporisation is not chemically congruent (i.e., the vapour phase has a bulk composition distinct from that of the residue; Markova et al., 1986; Sossi et al., 2019; Wolf et al., 2022; Fegley Jr et al., 2023). In general, volatilities of major rock-forming elements evaporating from silicate melts decrease in the order K > Na > Fe > Si > Mg > Ca > Ti > Al (Sossi et al., 2019). Therefore, the end-product of such a process would result in Ca-Al-rich silicate planets (Dorn et al., 2019). Additional models are needed to confirm whether the temperatures required for sustained loss of rocky material (with high molar mass) are likely too high, considering the large masses of super-Earths, to have resulted in significant loss (Benedikt et al., 2020; Erkaev et al., 2023).

### 3.3 Life-essential elements (CHNOPS) in rocky exoplanet mantles

The storage in planetary mantles of carbon, hydrogen, nitrogen, oxygen, phosphorous and sulfur is of particular interest because of their being universally required in known biology (see Krijt et al., 2022, for a recent review of the origin of these elements in planetary systems). These elements also stand out from those discussed in previous sections in purely physical terms: many of them readily form gasses or fluids at low pressures (i.e., they are 'volatile' elements) and they are some of the last elements to condense from a cooling nebular gas (if indeed they condense at the planet forming region at all). Their volatility leads to these elements partitioning into atmospheres and to their potential loss early in a planet's history (Watson et al., 1981; Kasting and Pollack, 1983; Tian et al., 2009; Zahnle and Catling, 2017). Partly for this reason, and partly by definition, CHNPS may only have trace abundances in the solid mantles of rocky planets, at least for relatively-oxidised silicate planets[2]. Even at trace quantities however, the properties of the CHNOPS elements can lead to them having a profound impact on the physics and chemistry of a planet.

The propensity of C-H-N-O-S, especially, to be lost or only partially accreted during planet formation makes their inventories particularly hard to predict. These elements, P included now, may all also enter planetary metal cores in proportions often strongly dependent on pressure, temperature and oxygen fugacity (e.g., Rubie et al., 2011). As a result, for exoplanetary systems there is (presumably) significant inherent stochasticity in the mantle abundances of CHNOPS, even if the systemic abundance is well constrained from the stellar composition (e.g., Sossi et al., 2022; Krijt et al., 2022; Lock and Stewart, 2023, see section 2.2).

---

[2] Excessive amounts of accreted volatiles would form a thick liquid ocean/ice layer atop the rocky mantle, producing a planet with a measurably lower bulk density that is likely identifiable as non-rocky (e.g., Unterborn et al., 2018, see Figure 1).

In the absence of being able to predict accurate mantle abundances of CHNOPS, one solution for some elements (H and N in particular) is to instead consider the limits internal conditions place on their storage: specifically, how temperature and mineralogy limit the concentration of these elements that can dissolve in the mantle. Such storage limits allow a minimum diversity of elemental abundances across planets to be inferred based on properties that can be estimated. The unconstrained accretion, loss, and redistribution history of CHNOPS would then increase this diversity. We are therefore here interested in the mineral storage of CHNOPS elements in planetary mantles, and how the mineralogical variation discussed in section 3.1 may in and of itself affect their presence in exoplanets.

The major mineral phases of exoplanet mantles can hold a large part of a silicate mantle's CHNOPS inventory. They do this by incorporating CHNOPS in their crystal lattices as non-stoichiometric substitutions for similarly-sized and -charged elements. Elements stored in this way have inherent solubilities in host minerals, above which the mantle would 'saturate' in the element concerned and need to stabilise a new mineral (or fluid) phase specifically incorporating that element to host any more. In the case where a solubility limit leads to a fluid being stabilised then, given the (upward) mobility of fluids, such solubility constraints in principle represent upper limits to the total mantle abundance of an element. CHNOPS elements do occur as saturated phases in Earth's mantle, in primitive meteorites, and likely in exoplanets' mantles by extension: C as diamond or graphite ($C^0$), or in carbides ($C^{4-}$) or carbonates ($CO_3^{2-}$), for example, and S in sulfides ($S^{2-}$) or sulfates ($S^{6+}$); here the form of either element will be a strong function of the local $fO_2$, because of different inherent redox states in these phases. Subsequent partitioning of CHNOPS from solids to silicate melts to magmatic gases will depend further on $fO_2$ and pressure (see Gaillard et al., 2021, for a recent review).

The redox-dependence of CHNOPS storage is another variable to factor into predictions of their abundance and mobility in a planet. As we are far from constraining an exoplanet's mantle $fO_2$ profile without knowing its detailed O budget, this is an additional source of uncertainty. Nonetheless, a useful constraint can be applied: a *minimum* variability of $fO_2$—of several log-units at constant $Fe^{2+}/Fe^{3+}$—emerges as a function of silicate mineralogy variability itself (section 3.1), via composition-dependent Fe activity coefficients (Guimond et al., 2023a).

By combining insights from mineralogically controlled solubility limits and $fO_2$'s, the mantle major element compositions of exoplanets can be linked to their maximum capacity for and speciation of CHNOPS. We briefly review storage of CHNOPS in this context below.

### 3.3.1 Carbon

The behaviour of carbon is tightly coupled to redox processes (Dasgupta, 2013; Stagno et al., 2019), and therefore the storage of carbon in exoplanetary mantles will be as diverse as their redox states. At low mantle carbon concentrations carbon will dissolve in major silicate minerals (e.g., olivine, orthopyroxene, clinopyroxene, and garnet), with solubilities on the order of 0.1–1ppm (Keppler et al., 2003; Rosenthal et al., 2015), yet below experimental detection limits in lower mantle minerals (Shcheka et al., 2006).

Carbon in excess leads to saturated accessory phases: carbides, graphite/diamond, and carbonates; the main carbon hosts in the terrestrial mantle, which for context may contain $\sim$10–100ppm C (Luth et al., 1999; Dasgupta and Hirschmann, 2010). The carbon phase diagram is a function of pressure, temperature, and importantly, of $f O_2$, as these phases' C oxidation states range from 4– to 4+. Which of these phases are actually stable in a mantle depends on where the $f O_2$ profile intersects the carbon phase diagram. We might generally expect $f O_2$ to decrease with depth because $Fe^{2+}$–$Fe^{3+}$ equilibria depend on pressure (Frost and McCammon, 2008). However, carbon itself may be a relevant redox buffer if it is abundant enough to overwhelm the Fe redox buffer, pointing to feedbacks between planetary C inventory and its speciation, and making predictions about exoplanet mantles complex.

Carbon cycles on exoplanets have nevertheless seen a number of recent modelling studies, picking up on the centrality of the silicate weathering stabilising climate feedback to the concept of the circumstellar habitable zone (Kasting et al., 1993). Yet no studies explicitly consider variable carbon mineral speciation in the mobility of carbon during magmatism and any related mantle outgassing. The solubility model of carbon in reduced basaltic melt from Holloway et al. (1992), assuming carbon only as graphite in the mantle magma source region, has appeared in several outgassing models (Hirschmann and Withers, 2008; Tosi et al., 2017; Honing et al., 2019; Ortenzi et al., 2020; Wogan̈ et al., 2020; Guimond et al., 2021; Baumeister et al., 2023). Here, carbon melting is controlled by redox reactions that move graphite into a dissolved carbonate component of silicate melt—melt carbon contents are therefore limited by $f O_2$. Meanwhile, other models (Foley and Smye, 2018; Unterborn et al., 2022) employ a constant carbon partitioning coefficient (e.g., Hauri et al., 2006) between silicate minerals and melt; i.e., assuming carbon at lower concentrations where it can be accommodated as a dissolved trace element in the major silicate minerals. Lastly, in the case of more oxidised mantles with more than a few ppm C, carbonate minerals would occur. Even low carbonate modalities ($\sim$10ppm) can suppress mantle melting temperatures significantly (leading to production of melts containing up to $\sim$half $CO_2$ by mass; Dasgupta and Hirschmann, 2006, 2010).

### 3.3.2 Hydrogen (and water)

H is relatively soluble in silicate crystals, minerals that are otherwise nominally anhydrous; dissolved H in common mantle minerals is a major reservoir of "water" in the mantles of mostly-solid rocky planets. Upper mantle silicates tend to have a water saturation limit on the order of 10–100 ppm, generally decreasing with temperature and increasing with pressure (e.g., Kohlstedt et al., 1996; Keppler and Bolfan-Casanova, 2006; Mosenfelder et al., 2006; Ferot and Bolfan-Casanova, 2010, 2012; Ardia et al., 2012; Tenner et al., 2012; Novella et al., 2014; Demouchy et al., 2017; Dong et al., 2021b; Andrault and Bolfan-Casanova, 2022)—with the notable exception of the transition zone minerals wadsleyite and ringwoodite holding wt.% water concentrations over a narrow annulus of pressure (e.g., Kohlstedt et al., 1996; Fei and Katsura, 2020; Bolfan-Casanova et al., 2023). At higher pressures, the water solubilities of nominally anhydrous perovskites, ferropericlase, and postperovskite are much more uncertain due to the technical difficulty of experiments at these conditions, but so far appear drier than olivine and orthopyroxene at the pressures investigated (Bolfan-Casanova et al., 2002, 2003; Litasov et al., 2003; Townsend et al., 2016; Chen et al., 2020; Liu et al., 2021; Ishii et al., 2022; Shim et al., 2022, cf. Murakami et al. 2002). Other factors such as $fO_2$ and C content may have secondary effects on the water capacities of these minerals (e.g., Bolfan-Casanova et al., 2023), though they tend to be neglected in studies of whole-mantle trends.

Integrating these water saturation limits across the expected range of silicate mineralogies and assuming a perovskite water capacity of 30ppm (Liu et al., 2021) leads to solid mantle water capacities ranging from ~250–2000ppm, depending mostly on mantle thickness, potential temperature, and Mg/Si ratio (Guimond et al., 2023b). These capacities could be tenfold higher if perovskite holds 1000ppm water instead (Inoue et al., 2010; Shah et al., 2021). Regardless, because the water capacities of various silicates stable at $\lesssim$ 15GPa are roughly similar, many rocky exoplanets may have accordingly similar maximum water capacities in the source region of their magmas (Guimond et al., 2023b).

The stabilisation of minerals with stoichiometric water in their structure would increase the water capacity of a mantle substantially. High-pressure hydrous silicate phases carrying up to ~10 wt.% water as a stoichiometric component, such as phases A, B, D, $\delta$, E, egg, superhydrous phase B, and $\delta$-H solution, may also exist in Earth's and exoplanetary mantles, but generally require low temperatures; i.e., geotherms that on Earth are associated with subducting slabs and therefore plate tectonics (Ringwood and Major, 1967; Liu, 1987; Kanzaki, 1991; Pacalo and Parise, 1992; Ohtani et al., 1997; Ohtani, 2021). Lastly, water appears fully miscible in silicate melts from ~4GPa (Bureau and Keppler, 1999; Mibe et al., 2007; Mookherjee et al., 2008). Melt trapped in mineral inclusions or along grain boundaries during rapid magma ocean crystallisation may retain interior water early in a planet's history (Hier-Majumder and Hirschmann, 2017). If a basal magma ocean endures, it would have huge potential as a water reservoir due to virtually unlimited water solubility (Moore et al., 2023; Boley et al., 2023). However, it is unclear how much this deep reservoir would contribute to subsequent water transport—whether it would leak to the lower mantle and eventually reach the upper mantle—especially if H diffusion in postperovskite is sluggish (Peng and Deng, 2023).

### 3.3.3 N–P–S

Nitrogen is soluble in mantle minerals as $NH_4^+$, substituting for $K^+$ (Hall, 1999; Yokochi et al., 2009; Watenphul et al., 2010; Li et al., 2013; Johnson and Goldblatt, 2015). Experiments by Li et al. (2013) found maximum $NH_4^+$ solubilities in enstatite orthopyroxene, diopside clinopyroxene, and pyrope garnet less than 100ppm (and much lower in forsterite olivine); solubilities markedly increase with decreasing $fO_2$ and increasing pressure. At sufficiently-reducing conditions, nitrides or N-Fe-alloys may occur (Daviau and Lee, 2021). Hence—and because lower $fO_2$ also favours more N in minerals over melt and gas phases (Libourel et al., 2003; Li and Keppler, 2014; Mikhail and Sverjensky, 2014; Dasgupta et al., 2022)—rocky planets with reducing magma oceans and mantles might retain significant fractions of primordial nitrogen dissolved in their silicates.

Sulphur and phosphorous, like carbon, will occur in planetary mantles both in accessory phases—sulfides or sulfates for S, and phosphides or phosphates for P, depending on $fO_2$ and (e.g., von Gehlen, 1992; O'Reilly and Griffin, 2000; Truong and Lunine, 2021)—and dissolved in major silicates when they are at lower concentrations. Solubilities in olivine are measured to be ~1ppm for S (but higher in the generally less-abundant clinopyroxene) and ~1000ppm for P (Brunet and Chazot, 2001; Callegaro et al., 2020). Thus, the main (terrestrial) mantle P host is expected to be olivine (Walton et al., 2021b), also supported by findings of P associated with olivine in analogous meteorites (e.g., Davis and Olsen, 1991; Walton et al., 2021a). Large differences in mantle S contents are suggested between Mercury, Earth, and Mars, possibly related to $fO_2$ differences (e.g., Gaillard and Scaillet, 2009; Malavergne et al., 2010; Nittler et al., 2011; Franz et al., 2019). The variability of S contents between rocky exoplanet mantles may have important implications for observable features of their atmospheres, as volcanism could lead to significant transient increases in their abundance of spectrally active sulfur species (Kaltenegger and Sasselov, 2010; Ostberg et al., 2023; Claringbold et al., 2023). The transport of P from planetary mantles is likely to have less direct atmospheric impacts (although the discussion of possible abiotic sources of phosphine on Venus is relevant here; Bains et al., 2022); however, it is indirectly critical for the possible presence of biosignatures, given phosphorus's universal requirement by known biology (See Walton et al., 2021b, for a recent review). Despite this lack of direct astronomical observables, phosphorus transport to the surface will readily occur via magmatism given its tendency to partition into magmas during partial melting of the mantle (Brunet and Chazot, 2001). In this way, the supply of P to hypothetical biospheres is regulated by the mantle inventory and geodynamics of the planet.

## 4 Melts and crusts of rocky exoplanets

The preceding discussion of CHNOPS elements in exoplanets emphasises not only their storage in mantles, but their transport out of them: it is through this connection between interior and surface/atmosphere that exoplanet mantles may impact habitability, climate, and atmospheric composition. Where this connection is strong, and surfaces/atmospheres are significantly perturbed physically and chemically by this mass exchange, it may ultimately be possible to infer interior properties and states from exoplanet observations, e.g., through volcanically produced atmospheres (Liggins et al., 2022) or surface reflectance spectra (Hu et al., 2012).

The connection between rocky exoplanet mantles and their surfaces occurs through partial melting, the process whereby changes in the pressure of the solid mineral assemblage of a mantle causes minerals to react (partially 'break down') to form a liquid. Melting may occur at various depths within mantles: melt may have potentially been left over at the base of the mantle from magma ocean solidification (Labrosse et al., 2007); melting may occur at the mantle transition zone where wet material from the upper mantle must lose its water into water-rich melts as it descends into the mineralogically dry lower mantle (Karato et al., 2020); or near the surface at low pressures where the mantle adiabat intersects the solidus (the locus of pressure-temperatures where solid mantle first begins to melt). It is this last region of melting that we will focus on here as it is the most directly connected to surface: melts from these low pressures feed almost all volcanism on Earth, whether at mid-ocean ridges, ocean islands like Hawaii and Iceland, or arc volcanoes. Adiabatic decompression melting, as this process is called, is also the mechanism of melt generation that is least contingent on the geodynamic mode of the planet. Whereas basal magma oceans may require a particular planetary composition and thermal evolution and wet melting at the transition zone may be predicated on plate tectonics, adiabatic decompression melting can take place in any convecting mantle. The youthfulness of volcanism on the small and cold Mars (Werner, 2009), on the dry and hot Venus (Herrick and Hensley, 2023), and plate tectonic Earth indicates the general role of adiabatic decompression melting in the dynamic homeostasis of terrestrial planets across a range of conditions and compositions. We can reasonably expect this type of melting to be a universal phenomenon on planets lacking thick volatile envelopes (i.e., rocky planets; Kite et al., 2009, 2016).

Partial melts of the mantle will move to low pressure if they are positively buoyant. This condition is met for melting in Earth's shallow upper mantle (and evidently that of Mars and Venus as well), leading to the planets' surface, 'crust', being built of the minerals these melts form on cooling. Even on Earth, where continental crust has been subsequently formed not by direct melting of the mantle, the majority of crustal surface area is ultimately the product of basalts derived from mantle melting. The mineralogy and composition of a planetary mantle is therefore closely connected to its crust, the substrate on which climatic and biological occur.

In this section we explore how the diversity of exoplanet mantles seen above will map through to their melt compositions and crusts. As discussed previously, the thermodynamic models that can calculate mineralogy are necessarily most accurate when applied to Earth-like compositions, yet here we are investigating the non-Earth-like compositions of diverse exoplanetary mantles. This uncertainty is likely magnified when calculating the composition of melts these mantles generate: the thermodynamic complexity of melts is inherently challenging even in trying to model melting of Earth-like mantle compositions. Nonetheless, qualitative insights can be gained that speak to the universality of petrological thermodynamics, even under somewhat exotic circumstances and we emphasise these below.

## 4.1 Mantle melting temperature

The first and most important aspect of mantle mineralogy as pertains to melt generation is the temperature the mantle melts at, its solidus temperature. The simple presence of melt, even when ≪ 1%, is enough to drastically change the rheology of a mantle, lowering its viscosity and thereby enhancing convection and heat transport (e.g., Kohlstedt and Zimmerman, 1996). Hirschmann (2000) investigated the effect of composition on solidus temperature for a range of broadly Earth-like mantle compositions based on experiments. This made clear the importance of both the relative proportion of MgO to FeO and the alkali element (Na, K) content of mantle rocks for their solidus temperatures. As discussed previously, whilst the Mg and Fe content of a mantle can likely be predicted from the stellar composition with some accuracy, the abundance of the alkali elements will be harder to predict, as both Na and K are moderately volatile during planet formation (Lodders et al., 2009). As a result, we here consider both Na-free and Na-bearing mantles, with the inclusion of the latter qualitatively indicating the magnitude of solidus reduction that mantle alkali contents make possible. As in previous sections, the intent with the calculations presented is to provide an indication of the variability we might expect among exoplanetary mantles, rather than accurately predict their properties.

The range of solidus temperatures predicted for the mantles of rocky planets is given in Figure 8. Results are shown for Na-free (filled histograms) and Na-bearing (open histograms) mantles, for three different upper mantle pressures in each case (1, 2 and 3GPa), and between panels the fraction of Fe in the core is varied.

The first order result of these calculations is to reproduce the effect Hirschmann (2000) identified from experiments, that the alkali content of the planetary mantle is able to cause lowering of the solidus temperature of over 100˚C. The magnitude of this effect for any given planet will be contingent on its devolatilisation history (and that of its building blocks), emphasising the likely stochasticity in this key property of exoplanet mantles. The effect is also broadly consistent across pressure and mantle Fe contents. Pressure variations themselves simply track the solidus to higher temperatures, as the positive volume change on reaction of producing a liquid is thermodynamically penalised at higher pressures, requiring greater temperatures to stabilise it.

If we look at single histograms in Figure 8, we see the effect of systemic (stellar) variability alone on solidus temperature. For Na-free mantles, even the quite substantial chemical variation reported in section 2 translates to only modest changes in solidus temperature, of typically less than 100°C. This is consistent with the finding of Hirschmann (2000) that variations in MgO/FeO, whilst important for solidus temperature, have much less of an effect than the alkali abundance. It is also consistent with expectations from simple model petrological systems, where melting occurs at the single eutectic temperature, irrespective of composition. This breaks down in detail for systems of real geological complexity (as modelled here, where minerals are complex solid solutions that change with bulk composition), but is the essential reason why large changes in major element composition lead to only modest changes in solidus temperature. For Na-bearing mantles however, we see that consideration of alkalis is important even for within-population variation: i.e., the variability of Na contents between planets simply inherited from the systemic composition, ignoring stochastic devolatilisation, leads to very large variation in solidus temperatures (up to 200°C).

A final variable considered in Figure 8 is the fraction of iron that goes into the planet's core versus that which stays in the mantle as FeO. Less FeO-rich mantles lead to higher melting temperatures, by up to ~ 100°C in the extreme case of almost all iron (98%) going into the core. The redox conditions of core formation therefore have a similar magnitude of effect on solidus temperature to intrinsic compositional variability at fixed core iron fraction.

Considering variation in mantle alkali content, iron content, and broader major element compositional variability together, rocky exoplanet mantle melting temperatures may vary by over 300°C. This purely compositionally-driven change in melting behaviour of planetary mantles has important implications for geodynamics, given the stabilising role melting has in heat transport and thereby planetary thermal evolution (e.g., Nakagawa and Tackley, 2012; Driscoll and Bercovici, 2014).

## 4.2 Melt and crustal compositions

Melting is the key process whereby mantle chemistry is transported to the surface to create crusts, influence climate, and planetary habitability. Melt chemistry changes progressively during partial melting, meaning any given mantle composition can produce a wide range of melt compositions according to the depth at which melting takes place and the extent to which melting proceeds above the solidus (where melt is technically 0% of the mass of the system) to the liquidus (where the system becomes entirely molten; a temperature range of hundreds of K or more). To simply give a first indication of how planetary mantle composition might map through to melt composition, we pick a single

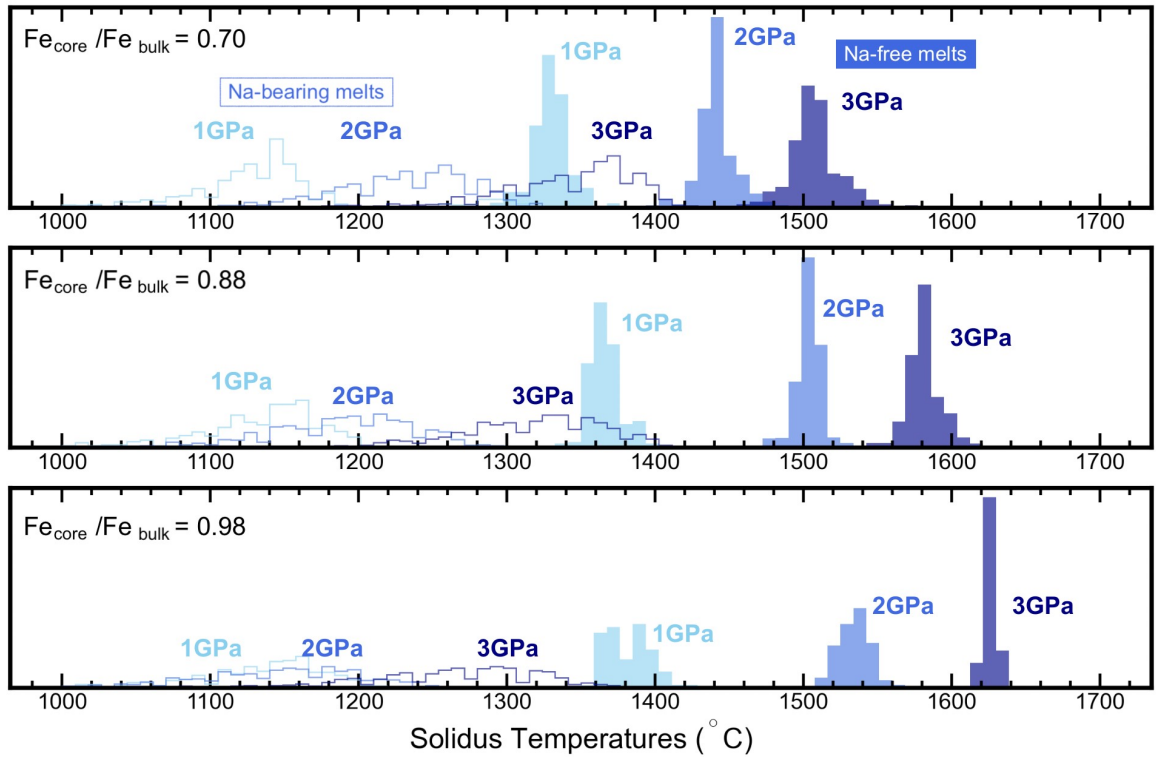

**Figure 8.** The solidus temperatures for sodium-bearing (open histograms) and sodium-absent (filled histograms) exoplanet mantles, calculated using pMELTS (Asimow and Ghiorso, 1998; Ghiorso et al., 2002). Mantle compositions are the same as in Figure 6, drawn from the Hypatia Catalog (Hinkel et al., 2014), but now also including a set of mantles containing Na to show the effect of minor elements on solidus temperature. Mantle Na abundances are calculated assuming a constant star/planet depletion factor equal to that between the sun and Earth. The effect of mantle FeO content on solidus temperature is also shown, with panels from top to bottom recording progressively less partitioning of bulk planet Fe to the mantle (constant molar ratios $Fe_{core}/Fe_{bulk}$ of 0.7, 0.8, and 0.98 respectively). degree of melting, 20%, and show results from this alone. This degree of melting has a rationale on Earth, as for typical Earth mantle it is at this degree that the mineral clinopyroxene is lost from the melting assemblage, beyond which point melting is less productive (i.e., efficient melting stops at this point; e.g., Katz et al., 2003). Such a petrologically informed melt fraction could be identified for all exoplanetary mantles, but given the assumptions already inherent to these calculations would be beyond the scope of this review.

The mapping between mantle composition and their partial melts is shown in Figure 9. The domination of silicate melts by a few oxides, $SiO_2$ and MgO in particular, makes the representation of the results as oxide versus oxide in weight percent hard to evaluate. However, even viewed this way, it is clear that a systematic trend is that melts are less magnesian than their corresponding mantles and have broadly similar silica contents. Transformation of the compositions to centred log ratios (emphasising relative variation, see caption to Figure 9) allows variation to be seen more clearly, particularly of the minor elements. Elements such as Na and Ti favor entering into the melt phase and can be seen to be systematically more abundant in melts than their corresponding mantle sources. CaO is also higher in melts than in mantles (due to its incorporation into clinopyroxene, which readily melts out), which is significant given the key role crustal Ca has in planetary carbon cycles via carbonate formation (e.g., Siever, 1968).

Another way of viewing the mapping of mantle to melt composition is by comparing the standard deviations of the respective populations: how does the variation in mantle composition (arising from stellar variability) compare to the variation that emerges once melting has taken place? Figure 10 shows this for melting calculated at three shallow upper mantle pressures. In all cases variation in $SiO_2$ and MgO in the melts is less than that in their sources, however for all other elements the absolute variation is greater. What this illustrates is that when it comes to creating planetary crusts, petrological thermodynamics can overprint the inherited bulk compositional characteristics of the body making prediction of the composition of exoplanetary crusts, highly non-deterministic.

Despite the high *compositional* variance, however, to a first order the mineralogies of these crusts are very similar: on cooling, these melts crystallise a similar mineral assemblage to that which they melted from in the mantle, dominantly olivine and clinopyroxene, with plagioclase feldspar incorporating most of the Al. Therefore, the compositional diversity seen in Figures 9 and 10 does not in and of itself propagate through to predictions of exoplanets having particularly exotic silicate crustal mineralogies. As Earth has demonstrated in forming a continental crust, subsequent differentiation processes—perhaps related to tectonic mode(s)—can transform the mineralogy and composition of primary igneous crusts dramatically.

### 4.3 Melts, crusts and geodynamics

The crustal composition of an exoplanet will have many consequences for its evolution, although many are too contingent for useful predictions to be made. One crustal composition consequence recently subject to deterministic modelling, however, relates to the fate of liquid water on the planet surface; something with wide reaching implications for habitability and geodynamics. Water may be consumed in reactions with crustal minerals, notably the serpentinization reaction whereby ferrous iron from mafic rocks dissolves in water to produce serpentine and $H_2$ gas—the resulting hydrated minerals can store significant weight fractions of $H_2O$. Certain crustal compositions may be more readily hydrated and hence destabilise surface liquid water (Herbort et al., 2020).

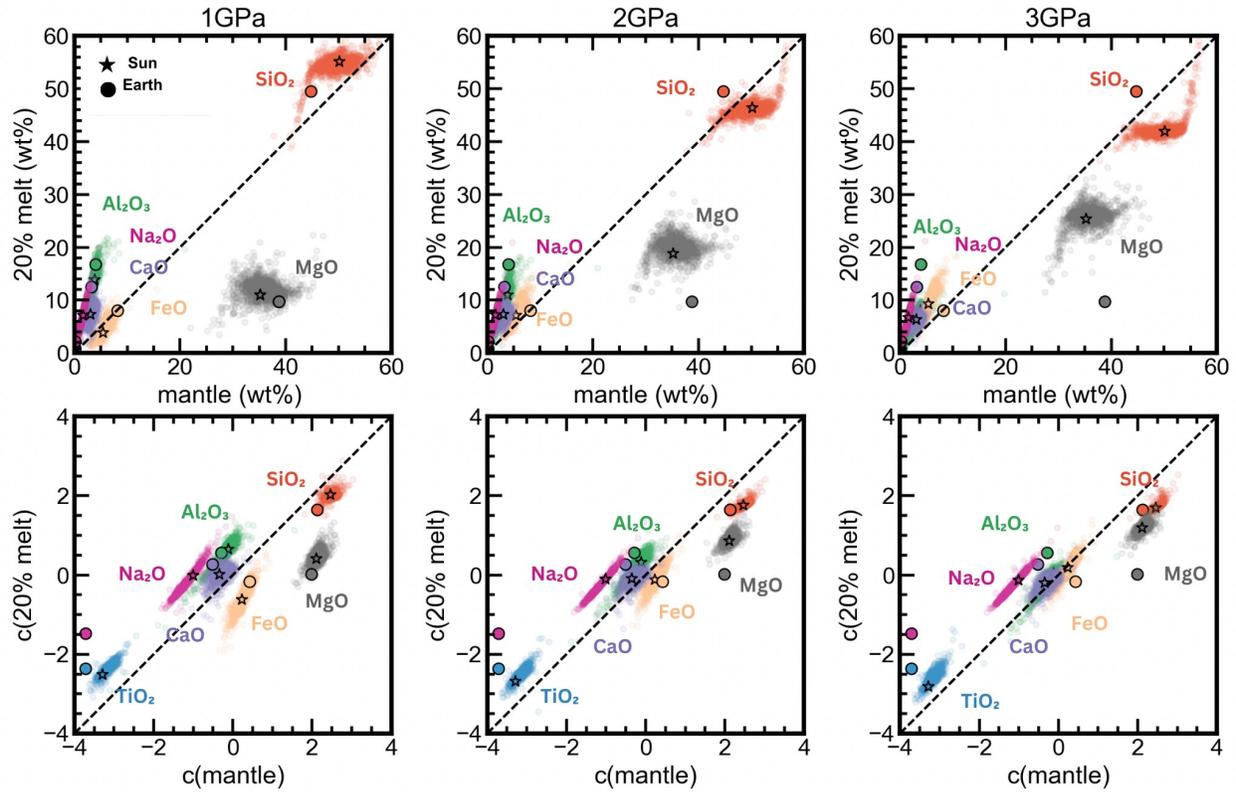

**Figure 9.** The diversity of melt compositions produced by 20% melting of rocky planet mantle compositions at a constant pressure, inferred from the Hypatia Catalog of stellar abundances and calculated using pMELTS as in Figure 8. Panels show melt compositions represented both as oxide weight percent *(top row)* and as centred log ratios *(bottom row;* 'c' denoting calculation of the centred log ratio from the weight percent oxide concentration). Centered log ratios preserve information on relative abundance of an element compared to the geometric mean and enable minor element variation to be seen more clearly (see Lipp et al., 2020, for a recent example of centred log ratios used and explained on geochemical data). Horizontal axes record mantle compositions; vertical axes record the compositions of their corresponding 20% melts. Equivalent calculations using the Earth's observed mantle composition (Workman and Hart, 2005, filled circle) and a solar-composition mantle (Lodders et al., 2009, filled star) are shown for reference.

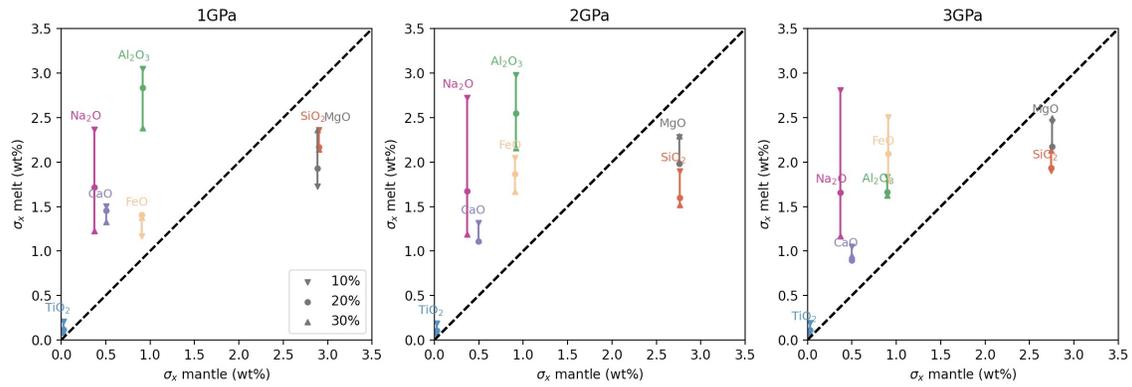

**Figure 10.** The variability of rock-forming oxides in mantle-derived melts compared to the variability of the same oxides in mantles, as measured by the standard deviation of their oxide weight proportions ($\sigma_x$). Data and modelling are the same as in Figures 8 and 9. Panels show calculations at three different melting pressures (1, 2, and 3 GPa; *left to right*), with symbols indicating different assumed melt fractions (10%, 20%, and 30%). Dashed lines indicate the 1:1 correlation between mantle $\sigma_x$ and melt $\sigma_x$.

Notably, increasing the FeO content of basalt stabilises these hydrated minerals (due in part to smaller volume changes during the hydration reaction), and hence on the crust's ability to sequester surface water (Wade et al., 2017; Dyck et al., 2021). On Mars, which is relatively FeO-rich but otherwise similar in composition to BSE, past serpentinization could have removed hundreds of metres of an equivalent global ocean to the crust, and eventually to the mantle, given enhanced (hydrous) amphibole stability (Wade et al., 2017). Indeed, phyllosilicates (hydrously altered crustal minerals) are observed widely on the martian surface (Lammer et al., 2013; Carter et al., 2013), and Martian D/H measurements support $\sim$ 30–99% of early surface water now residing in the crust (Scheller et al., 2021). This result points to a direct effect of core formation on the surface environment of a rocky planet.

The potential fate of water on a planet emphasises that crusts act as a second silicate reservoir on a planet, and potentially a major one depending on the element concerned: particularly the 'incompatible' elements that strongly enter the melt phase, e.g., Ti and heat producing elements U, Th, and K, and elements sequestered into the crust by low temperature processes, e.g., C during formation of carbonate minerals, which helps regulate long term climate. The role of the crust in geodynamics is therefore sensitive to the amount of mantle that has been processed by melting to produced it (i.e., has the whole mantle lost its K in producing the crust, or just a small fraction) and its longevity (i.e., is the crust routinely recycled back into the mantle or is it a permanent reservoir). Both of these points indicate the strong feedbacks that will emerge on a planet between crust production/destruction and its surface and interior dynamics. This history of these processes on Earth has been hotly debated for many decades (e.g., Korenaga, 2018, for a recent review). Rather than being tightly constrained in how these processes may operate, exoplanetary science ultimately has the opportunity to resolve some of these debates with new observations of rocky planet evolution.

## 5 Observing exoplanet compositions and mineralogies

As discussed in the previous sections, theoretical attempts to predict the mantle mineralogy of rocky exoplanets are widespread but untested. Yet a major opportunity presented by exoplanets is empirical as well as theoretical: to extrapolate knowledge from Earth science by applying it to strange planets, but also to rewrite this knowledge using unprecedented observations. The statistical leverage of exoplanet demographics (i.e., planet mass and period distributions) has already re-hauled theories of planet formation, for example, previously only constrained by one planetary system (Mordasini, 2018). Since we have seen that mass and radius are inherently limited in what they constrain about a planet's composition, even with the host star as a proxy (e.g., Sotin et al., 2007; Grasset et al., 2009; Rogers and Seager, 2010; Santos et al., 2015; Dorn et al., 2015; Hinkel and Unterborn, 2018; Wang et al., 2019a; Schulze et al., 2021), observations beyond mass and radius (important and hard won as they are) are needed to reveal to us the geological complexities of rocky planets.

In this section we describe some of the methods of observation that could help constrain exoplanets' mantle, melt, and crust compositions. Two of the main astrophysical signals carrying *chemical* information on exoplanets are: *(i)* the proportion of a planet-hosting star's photons, detected by a telescope, which have been absorbed by a planet's atmosphere as the planet passes in front of ("transits") that star—this technique is called transmission spectroscopy if the photon wavelengths are measured—and *(ii)* when the planet is instead behind the star ("in eclipse"), the proportion of detected photons now blocked by the star, which would be either emitted or reflected by the planet— this is called emission spectroscopy, and the planet's now absent contribution is identified by difference compared with when it is present just prior to entering into eclipse. A quantity often reported in these measurements is the flux ratio of photons, often measured in ppm, between the star during planetary eclipse or transit, and the star alone. In general, higher planet/star flux ratios permit better measurement precision. The James Webb Space Telescope (JWST), for example, is able to measure this ratio as a function of wavelength to unprecedented precision. Hence many of these observations have only recently become feasible given the already-proven success of JWST since its launch in 2021.

## 5.1 Lava worlds as a window into rocky planetary interiors

Among the potentially rocky exoplanets discovered thus far, there is a particularly interesting class that could enable observational constraints on exoplanet silicate compositions: magma ocean planets (or lava worlds). These worlds, of which no analogue exists in our present solar system, orbit their host stars on very short orbits, usually less than a few days. Such proximity to their stars implies scorching dayside temperatures hot enough to melt mantle rock at least to some depth. So far, a couple hundred such objects have been identified as candidates in transit surveys (Zilinskas et al., 2022), and a few dozen of them have mass measurements as well (see Figure 1). Of this subset, the corresponding average densities strongly hint at silicate interiors, and consequently, silicate-dominated surfaces.

Ultra-hot lava ocean planets are typically expected around FGK-stars, as only these are bright enough to raise temperatures of close-in planets sufficiently for extensive melting. Given the relatively high intrinsic (thermal) emission of a hot lava planet, the planet-to-star photon flux ratio can reach values up to 300 ppm,[3] but are typically expected to be $\sim$ 100 ppm in the mid-infrared (for a 2$R_E$ Super-Earth orbiting a Sun-like star; Zilinskas et al. 2022). Additionally, their short orbital periods ($\leq$ 2 days) imply that large numbers of orbits can be observed, resulting in improved statistics. This means that lava planets are our only opportunity to study the surfaces of rocky planets around Sun-like stars with contemporary facilities.

The intense temperatures these planets experience imply that lava oceans cover large (but not necessarily deep) surface areas of their starlit hemispheres (Leger et al., 2011). That is, planets on such close-in orbits likely have a´ permanently-irradiated hemisphere due to being tidally locked (as the moon is to Earth; e.g., Barnes, 2017). Hence they would show strong hemispherical asymmetry in their surface conditions, with the nightside potentially being solid subject to the efficiency of heat transport through any atmosphere. Current classification of lava planets assumes a rule of thumb that surface temperatures must be in excess of ~1600K in order to melt rock (Kite et al., 2016, see also Figure 8). Since the main source of heat is radiation from the star, the lava ocean would be heated from the top rather than from the bottom, implying strong thermal (and potentially chemical) vertical stratification (Boukare et al., 2022,˙ 2023; Meier et al., 2023). In this configuration, the ocean would preferentially convect laterally, from the substellar point (where the star is directly overhead) to the terminator (where the star disappears over the horizon and the nightside begins). However, the associated heat transport is likely too inefficient to modify the surface temperature distribution by magma convection alone, meaning that, barring heat transport in an atmosphere, the hottest place on the planet's surface is still expected to be the substellar point (Kite et al., 2016).

---

[3] In Zilinskas et al. (2022), larger flux ratios of up to 500 ppm are reported for TOI-1807 b, based on an overestimation of the radius. However, its radius was overestimated and has been corrected since (Hedges et al., 2022).

In any case, at these hot dayside temperatures, species in the silicate melt will evaporate from the magma ocean to produce an atmosphere. The presence of this silicate vapour atmosphere is what opens up the study of lava worlds' interior compositions via emission spectroscopy (the hot atmosphere confined to the dayside implies that lava worlds have to be observed in eclipse, when the planet's starlight hemisphere is pointing towards the observer). Here, the pressures reached by rock vapour alone can be remarkable especially at the highest temperatures: whilst they are tenuous ($P \leq 10^{-5}$ bar) at $T < 2000$ K), up to Mars-like pressures are expected at $T \sim 3000$ K (Fegley and Cameron, 1987; Wolf et al., 2022; Van Buchem et al., 2022). Chemically, models predict these atmospheres to be dominated by Na, K and Fe gas at low temperature, and SiO at higher temperatures (Miguel et al., 2011; Zilinskas et al., 2022; Wolf et al., 2022; Van Buchem et al., 2022).

Yet because the exact gas composition depends on the chemistry of the underlying melt, the observable atmosphere is linked directly to the planets' silicate compositions. Inverting an atmospheric composition from observations should therefore allow us, in principle, to constrain a magma ocean surface composition—e.g., bulk oxide abundance (Zilinskas et al., 2022) and oxygen fugacity (Wolf et al., 2022; Sossi et al., 2019)—and by extension, potentially an underlying mantle composition as well, since the two will be linked through the process of partial melting (see Sec. 4).

A key feature of these atmospheres is their confinement to the partially-molten dayside. This is because a certain vapour pressure above the melt is needed to sustain the atmosphere, so without any surface melt, the atmosphere will condense and collapse. Preliminary calculations involving simple hydrodynamics—i.e., winds directed from the substellar point towards the nightside—show that atmospheres cannot extend beyond the day-night terminator, and will have the bulk of their mass contained within a few tens of degrees from the substellar point (Kite et al., 2016; Nguyen et al., 2022). Beyond this point, condensation from gas to solid/liquid will assure atmospheric collapse.

Inverting these observations for atmospheric composition is nevertheless complex, in part because the inversion relies on understanding the atmospheric physics involved. For example, vertical 1D radiative transfer models have hinted at pronounced thermal inversions, a result of the strong absorption by SiO in the UV. This would mean that the atmosphere exhibits emission features in the infrared. In particular, SiO would show broad emission features, allowing it to be identified (Ito et al., 2015; Zilinskas et al., 2022). Initial efforts to observe these worlds have already been made. Demory (2014), using eclipse observations with the Kepler space telescope, concluded that some hot rocky exoplanets show high albedos—consistent with cloud coverage (Mahapatra et al., 2017; Mansfield et al., 2019) or wavy magma oceans (Modirrousta-Galian et al., 2021). Using the Spitzer space telescope to observe K2-141 b, an ultra-hot rocky exoplanet orbiting a K-dwarf star, Zieba et al. (2022) found their data to be consistent with no dayside-nightside heat redistribution—ruling out a thick atmosphere of a few bars, but as expected for both a thinner (≲0.1 bar) rock vapour atmosphere, and no atmosphere. The resolution of Spitzer was insufficient to detect any spectral features of SiO or other mineral species.

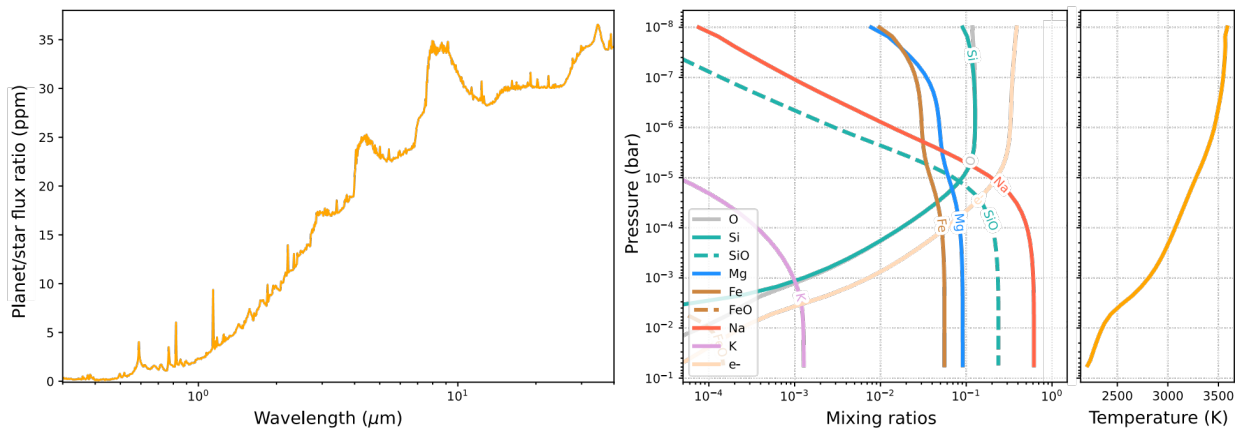

**Figure 11.** Properties of a silicate atmosphere vapourised from a Bulk Silicate Earth melt at 2500 K irradiation temperature from a solar-type star. *(Left:)* Spectrum of a lava planet of $1M_E$ and $1R_E$. *(Center:)* Atmospheric speciation, given in volume mixing ratio. *(Right:)* Temperature-pressure profile; a strong thermal inversion is pronounced in this atmosphere, leading to the emission features evident in the left plot.

Another ultra-hot planet studied extensively is 55 Cancri e, even though its nature as a purely rocky world might be challenged based on its low density (Dorn et al., 2017; Schulze et al., 2021). Initial considerations dismissed the presence of a primary atmosphere derived from the nebular gas, given the absence of significant H Lyman-$\alpha$ detection during transit observations with the Hubble Space Telescope (Ehrenreich et al., 2012), rendering a rock vapour atmosphere more likely. Despite persistent efforts, attempts to characterize this planet's atmosphere are yet to be successful. Studies by Ridden-Harper et al. (2016) and Tabernero et al. (2020) failed to detect Na (a moderately volatile element) and Ca (a refractory element). Similarly, Jindal et al. (2020) ruled out $H_2O$ and TiO based on cross-correlation analyses with high-resolution ground-based data. A comparable outcome was reported by Keles et al. (2022), whose exhaustive high-resolution search encompassed a broad spectrum of gaseous species, yet yielded no positive detections. However, all these observations were performed in transmission spectroscopy during primary transit, which is not the ideal way to detect an atmosphere composed entirely of rock vapour.

Meanwhile, there is some evidence that 55 Cnc e harbours a substantial atmosphere, or at least a transient one (Meier Valdes et al., 2023; Heng, 2023). Some studies suggest that there might be a shift in the surface's hottest point' away from the substellar point, indicative of an atmosphere that redistributes heat (Demory et al., 2016; Angelo and Hu, 2017; Hammond and Pierrehumbert, 2017). A reanalysis of the same data could not reproduce their findings, however (Mercier et al., 2022). 55 Cnc e was subject to three observing programs in Cycle 1 of JWST.[4] By the time of release of this article, there might be further information on this enigmatic ultra-short, massive super-Earth.

Discussion above was focused on "dry" lava planets, worlds that either formed with low volatile abundances, or lost volatiles through different physical drivers (cf. Owen and Wu, 2017; Ginzburg et al., 2018). If volatiles such as $H_2$, $H_2O$, CO or $CO_2$ are present in lava world atmospheres, they would significantly modify the spectral features of such atmospheres, and potentially wash out any signs of silicate vapour atmospheres (Piette et al., 2023). Further, the mere presence of volatiles significantly changes the chemistry of the melt and the vapour (Charnoz et al., 2023); mutual interactions modify both reservoirs (e.g., Schlichting and Young, 2022). This is an area where experimental petrology of magma degassing is critical in informing the design and interpretation of observations (e.g., Bower et al., 2019): just one reason why it is crucial to continue searching for atmospheric species with JWST, and future missions, in coordination with building our experimental and theoretical understanding.

**5.2 Dust tails of disintegrating planets**

Small, roughly Mercury-sized lava planets may not be able to retain their silicate atmospheres and could catastrophically evaporate. This process would leave a dusty outflow similar to a cometary tail, or even a torus along the orbit, which could be detected via its effects on the shape of the planet/star flux ratio as it varies in time over an orbit (Rappaport et al., 2012; Brogi et al., 2012; Budaj, 2013; Perez-Becker and Chiang, 2013; Booth et al., 2023). So far, three potential candidates for catastrophically evaporating planets have been identified on this basis: K2-22 b, KIC 1255b and KOI-2700 b. Comparing both spectral and albedo-versus-orbital phase observations of the planet to synthetic observations of known composition could constrain the composition of the dust (e.g., van Lieshout et al., 2014, 2016; Curry et al., 2023; Booth et al., 2023; Estrada et al., 2023)—major mineral species in these dust tails could be detected by JWST, for example (Okuya et al., 2020). The survival time of the dust might additionally provide information on its composition (e.g., Bromley and Chiang, 2023). Whilst no definite conclusions can be drawn from initial observations, Mg-Fe silicates (as opposed to pure Mg silicates or pure $Al_2O_3$) are a plausible match for KIC 1255 b and K2-22 b (Estrada et al., 2023).

---

[4] JWST Cycle 1 Proposals #1952, #2084, and #2347

**5.3 Reflectance spectra and surface mineralogy**

As emphasised in section 4, a planet's crustal composition and mantle composition are tightly related. If a rocky planet has lost its atmosphere, then its surface composition may be accessible through reflectance and emission spectra (Hu et al., 2012). Minerals have their own intrinsic emission and absorption spectra, which may be distinguished (e.g., Sunshine et al., 1990), or even be traced for the composition of individual minerals (e.g., Lucey, 1998). These methodologies have been thoroughly investigated in remote sensing on Earth and other objects of the Solar System to infer the composition of the surface (e.g., Lucey et al., 1998) or dust in Mars' thin atmosphere (e.g., Hamilton et al., 2005).

An exposed planetary surface will experience impact from micrometeorites, which will turn bulk rock into a fine porous material called regolith. Most exposed surfaces of atmosphere-less Solar System objects are covered with regolith. Due to its porosity, its reflective properties differ significantly from bulk rocks, as light is scattered multiple times between and within grains (Hapke, 2012). This creates a whitening effect, and may even conjure strong back-reflection at superior conjunction (Hapke et al., 1993; Hapke, 2002). Further, as the surface is subject to space weathering from energetic particles of the solar wind, its spectral features might change, alongside an overall darkening of the surface (Hapke, 1973). In principle, this could allow investigation of the age of a planet's surface—with a young surface implying recent volcanism.

Nevertheless, even in regolith, the mineral grains will imprint their spectral signature on the reflected (or emitted) light. Absorption features are usually broad, but unique to the mineral and its composition. $SiO_2$ shows a particularly strong feature at the 9 $\mu m$ wavelength band, which could be used to determine the silica content of (exo-)planetary surfaces (Hu et al., 2012; Kreidberg et al., 2019). Iron may also reveal its presence in the shift of absorption bands of olivine in the near infrared (Sunshine et al., 1990; Lucey, 1998). When minerals are not in their crystalline form, for example due to the formation of glasses (either through rapid cooling of extruding melt or impact breccia), the spectral signature may still be preserved, although it may differ from mineral spectra. This effect could help determine not only the composition, but also the material state of a planetary surface (Fortin et al., 2022).

Observations of rocky planet surfaces are relatively easier to conduct for hot planets ($\sim$ 500–1000 K) orbiting cool stars. In these cases, the planet-to-star flux ratio will be high. The surface of LHS 3844 b, a hot (1000 K) planet orbiting an M-dwarf, has been studied with the Spitzer infrared space telescope at 4.5 $\mu m$. In this case, no evidence of a thick atmosphere was found, enabling observers to believe their flux ratio measurement contained a signal direct from the planet's surface (Whittaker et al., 2022). Even with the single wavelength band observed, a surface composed of bright feldspathic or granitic material—common on Earth's continents and on the moon—could be ruled out tentatively, favouring a dark surface with a basaltic or ultramafic lithology (the primitive product of mantle melting, section 4, and as seen on Mars, for example). For the hot and also likely airless rocky planet GJ 1252 b, observations again with the Spitzer telescope suggested a similarly dark surface, but were less conclusive (Crossfield et al., 2022). Further studies of planetary surfaces with JWST (which, like Spitzer, operates in infrared wavelengths) have been proposed and accepted for a few of-interest targets [5], potentially delivering detailed results by the time this chapter has been published.

## 6 Summary

This review has focused on how we might predict the composition and mineralogy of exoplanet mantles and their resultant crusts. These predictions are rooted in knowledge of stellar composition, and the assumption that the composition of the star, being so much the mass of the system, is representative of the material available to build planets in the disk. As we have emphasised, deductions of composition and mineralogy become more uncertain the further we take them: the planet's composition may have been affected by devolatilisation/incomplete condensation from the nebula gas; the mantle composition is affected by an unknown core formation processes; the character of melts from that mantle are contingent on the depth and degree of melting, and so on. Whilst experiment and theory, often originating in the Earth sciences, make each step somewhat deterministic, the ensemble of branching points between a stellar composition and surface reflectance spectra, for example, make accurate prediction impossible. Nonetheless, we hope to have shown that petrological thermodynamics does introduce some certainty, even in the face of great diversity: upper and lower mantles separated by a transition zone are likely ubiquitous; and olivine and pyroxenes are likely to almost always be the major (solid) mantle silicate phases. More indirectly, the productive region of mantle melting, after incompatible minor element exhaustion, is likely to lie at a similar temperature across rocky exoplanets, and primitive crusts, derived from melting these mantles, will be broadly basaltic.

More importantly than the planetary properties we can predict, however, is that fact that our forward models of mantle mineralogy and melt chemistry are the prelude to inferring exoplanet geology from observations. Whilst detailed characterisation of rocky exoplanets is nascent, already we are probing hot planets for the compositions of any (possibly silicate) atmospheres. When such observations are secured, and the composition or mineralogy of an

---

exoplanet is even weakly constrained, we will have the first evaluation of how the assumptions in our predictive models have fared. This sparks the exciting era of testing and revising these models in light of exoplanet observations. We look forward to new insights into the physics and chemistry of rocky planets, their evolution, and their interior processes: insights equally as valuable for developing our understanding of Earth as for searching for habitable and inhabited exoplanets.

## Acknowledgements

The authors would first like to acknowledge the supreme patience and gracious and supportive handling of the manuscript by the Editorial team, and in particular an insightful review from Keith Putirka that helped improve the communication in the manuscript greatly. The research shown here acknowledges use of the Hypatia Catalog Database, an online compilation of stellar abundance data as described in Hinkel et al. (2014), which was supported by NASA's Nexus for Exoplanet System Science (NExSS) research coordination network and the Vanderbilt Initiative in Data-Intensive Astrophysics (VIDA). We also acknowledge use of the Third Data Release of the GALAH Survey. The GALAH Survey is based on data acquired through the Australian Astronomical Observatory, under programs: A/2013B/13 (The GALAH pilot survey); A/2014A/25, A/2015A/19, A2017A/18 (The GALAH survey phase 1); A2018A/18 (Open clusters with HERMES); A2019A/1 (Hierarchical star formation in Ori OB1); A2019A/15 (The GALAH survey phase 2); A/2015B/19, A/2016A/22, A/2016B/10, A/2017B/16, A/2018B/15 (The HERMES-TESS program); and A/2015A/3, A/2015B/1, A/2015B/19, A/2016A/22, A/2016B/12, A/2017A/14 (The HERMES K2-follow-up program). CMG was supported by the Science and Technology Facilities Council [grant number ST/W000903/1]. HW, FS and PS were supported by the Swiss National Science Foundation (SNSF) via an Eccellenza Professorship (203668) and the Swiss State Secretariat for Education, Research and Innovation (SERI) under contract number MB22.00033, a SERI-funded ERC Starting Grant '2ATMO' to PS.